\documentclass{emulateapj}
\shorttitle{SN~2006tf Gets the Bronze}
\shortauthors{Smith et al.}
\begin{document}

\title{SN~2006\lowercase{tf}: Precursor Eruptions and the Optically
Thick Regime of Extremely Luminous Type II\lowercase{n} Supernovae}

\author{Nathan Smith, Ryan Chornock, Weidong Li, Mohan Ganeshalingam,
Jeffrey M.\ Silverman, \\ Ryan J.\ Foley, \& Alexei V.\ Filippenko}
\affil{Department of Astronomy, University of California, Berkeley, 
CA 94720-3411; nathans@astro.berkeley.edu}

\and

\author{Aaron J.\ Barth}
\affil{Department of Physics and Astronomy, University of California,
4129 Frederick Reines Hall, Irvine, CA 92697-4575}

\begin{abstract}

  SN~2006tf is the third most luminous supernova (SN) discovered so
  far, after SN~2005ap and SN~2006gy.  SN~2006tf is valuable because
  it provides a link between two regimes: (1) luminous type IIn
  supernovae powered by emission directly from interaction with
  circumstellar material (CSM), and (2) the most extremely luminous
  SNe where the CSM interaction is so optically thick that energy must
  diffuse out from an opaque shocked shell. As SN~2006tf evolves, it
  slowly transitions from the second to the first regime as the clumpy
  shell becomes more porous.  This link suggests that the range in
  properties of the most luminous SNe is largely determined by the
  density and speed of hydrogen-rich material ejected shortly before
  they explode.  The total energy radiated by SN~2006tf was at least
  $7 \times 10^{50}$ ergs.  If the bulk of this luminosity came from
  the thermalization of shock kinetic energy, then the star needs to
  have ejected $\sim$18 M$_{\odot}$ in the 4--8 yr before core
  collapse, and another 2--6 M$_{\odot}$ in the decades before that.
  A Type Ia explosion is therefore excluded.  From the H$\alpha$
  emission-line profile, we derive a blast-wave speed of 2,000 km
  s$^{-1}$ that does not decelerate, and from the narrow P Cygni
  absorption from pre-shock gas we deduce that the progenitor's wind
  speed was $\sim$190 km s$^{-1}$.  This is reminiscent of the wind
  speeds of luminous blue variables (LBVs), but not of red supergiants
  or Wolf-Rayet stars.  We propose that like SN~2006gy, SN~2006tf
  marked the death of a very massive star that retained a hydrogen
  envelope until the end of its life, and suffered extreme LBV-like
  mass loss in the decades before it exploded.

\end{abstract}

\keywords{circumstellar matter --- stars: evolution --- stars: mass
  loss --- stars: winds, outflows --- supernovae: individual
  (SN~2006tf)}

\section{INTRODUCTION}

Massive stars that die as core-collapse supernovae (SNe) of Types II
and Ib/c span a wide range in luminosity, but they are typically
fainter than the standard thermonuclear Type Ia events that mark the
deaths of lower-mass stars.  More-luminous counterexamples are usually
observed as SNe of Type IIn, exhibiting relatively narrow ($\sim 1000$
km s$^{-1}$) emission lines of H in their spectra (Schlegel 1990; for a review
of SN classification, see Filippenko 1997).  These lines are generally
attributed to shock interaction with dense circumstellar matter (CSM)
rather than photospheric, high-velocity emission from the SN
ejecta. They can show multiple components: a very narrow feature
having typical widths up to a few hundred km s$^{-1}$ attributed to
emission from pre-shock CSM shed by the progenitor, as well as broader
components with widths of a few thousand km s$^{-1}$ arising from
dense post-shock gas.  We will refer to these two components as
``narrow'' and ``intermediate-width'' lines, respectively.  By
comparison, broad lines in SN ejecta typically have widths of
10,000--20,000 km s$^{-1}$.

Sufficiently dense CSM can decelerate the blast wave and convert its
bulk kinetic energy into X-rays and then visual radiation, thereby
substantially increasing the bolometric luminosity of the SN.  This is
seen in luminous SNe IIn and II-L having very strong H$\alpha$
emission, such as SNe~1979C, 1986J, 1988Z, 1994W, 1997cy, and others
(Branch et al.\ 1981; Filippenko 1991, 1997; Leibundgut et al.\ 1991;
Turatto et al.\ 1993; Sollerman et al.\ 1998; Chugai et al.\ 2004;
Germany et al.\ 2000; Turatto et al.\ 2000; Benetti et al.\ 1998), as
well as the ``hybrid'' Type Ia/IIn objects SN~2002ic (Hamuy et al.\
2003; Wang et al.\ 2004; Wood-Vasey et al.\ 2004; Kotak et al.\ 2004;
Benetti et al.\ 2006) and SN~2005gj (Prieto et al.\ 2007; Aldering et
al.\ 2006). SNe~1997cy and 1999E may also be hybrid SNe of this type
(Germany et al.\ 2000; Filippenko 2000; Rigon et al.\ 2003; 
Hamuy et al.\ 2003; Wood-Vasey et al.\ 2004).

\begin{figure*}
\epsscale{0.9}
\plotone{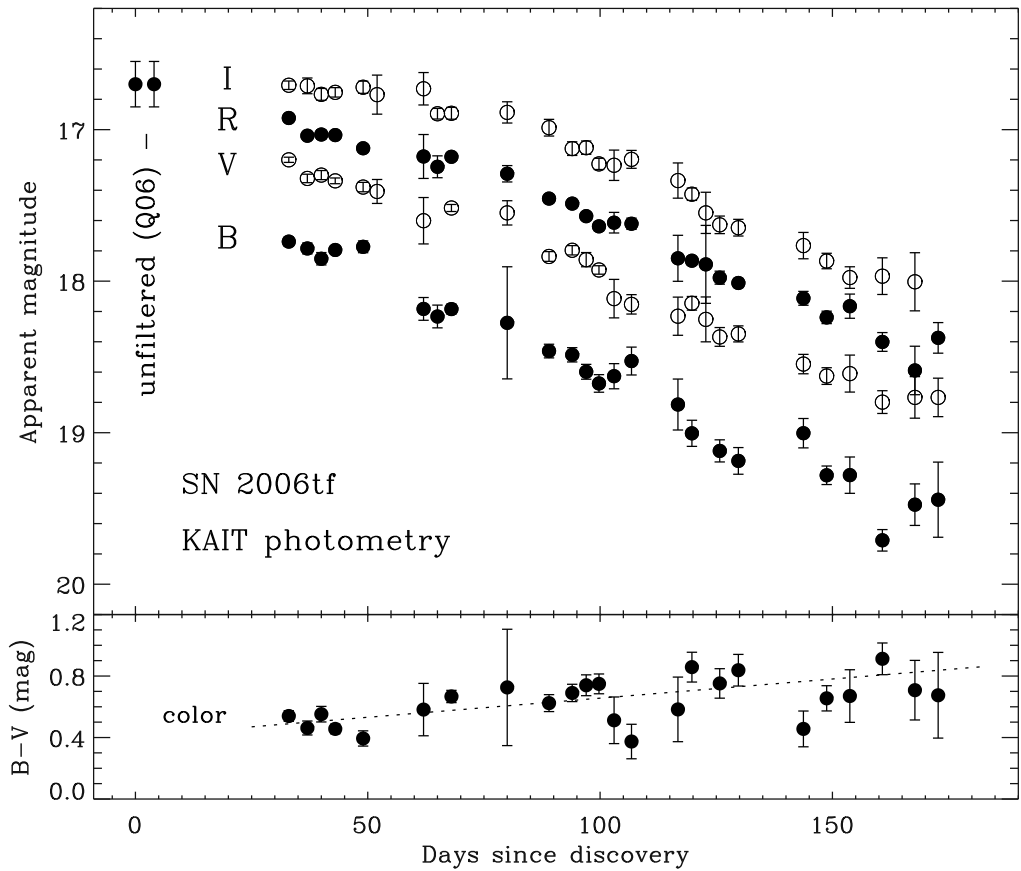}
\caption{The top panel shows our {\it BVRI} light curves of SN~2006tf
obtained with KAIT (see Table 1), and the bottom panel gives the apparent
$B-V$ color.  The dotted line illustrates a least-squares fit to the color
change with a slope of 0.0025 $\pm$0.0003 mag d$^{-1}$.  Days since
discovery are plotted because the explosion date is not
known.}\label{fig:mags}
\end{figure*}

Because it is the dense CSM that absorbs momentum and drains kinetic
energy from the blast wave, more-luminous SNe~IIn require progenitors
with higher mass-loss rates or slower wind speeds.  To account for
some of the more-luminous SNe~IIn, inferred progenitor mass-loss rates
need to be of order 0.1~M$_{\odot}$ yr$^{-1}$ or higher (e.g., Chugai et
al.\ 2004; Chugai \& Danziger 2003; Smith 2008).  These extreme
requirements point to episodic mass ejection reminiscent of the
eruptions seen in $\eta$ Carinae and other luminous blue variables
(LBVs), which have mass-loss rates of 0.01--1 M$_{\odot}$ yr$^{-1}$,
far exceeding the limiting mass loss in line-driven stellar winds
(Smith \& Owocki 2006; Owocki et al.\ 2004).  In SN~IIn progenitors,
the reason for this extreme mass loss is not yet understood, but it
must occur in the decades preceding core collapse because the blast
wave can only reach radii out to a few 10$^2$ AU during early phases
of these SNe.

Two recent events have pushed the limits on the required physical
conditions, challenging the standard picture of CSM interaction:
SN~2006gy (Ofek et al.\ 2007; Smith et al.\ 2007; Woosley et al.\
2007; Smith \& McCray 2007) radiated more energy in visual light than
any other known SN, and SN~2005ap (Quimby et al.\ 2007a) appears
to have had the brightest peak absolute magnitude yet observed.  To
account for their extreme luminosities with CSM interaction, the
required progenitor mass-loss rates are of order 1~M$_{\odot}$
yr$^{-1}$.  The problem is that in SN~2006gy, the weak H$\alpha$ and
X-ray emission imply mass-loss rates a factor of 10$^2$--10$^4$ smaller
(Smith et al.\ 2007), while SN~2005ap shows no spectral signature of
CSM interaction at all (i.e., it was {\it not} a SN~IIn; Quimby et
al.\ 2007a).  One possible solution is that these SNe are indeed
ultimately powered by CSM interaction, but this interaction occurs
early in a highly opaque shell, forcing photons to diffuse out {\it
after} the shock has already passed through it (Smith \& McCray 2007).
An alternative view (Smith et al.\ 2007; Quimby et al.\ 2007a) is that
these two SNe could be powered by radioactive decay from a large mass
of $^{56}$Ni synthesized in a pair-instability SN event (Barkat et
al.\ 1967; Rakavy \& Shaviv 1967; Bond, Arnett, \& Carr 1984), but
that hypothesis remains difficult to prove or rule out (see Smith et
al.\ 2008b), partly because the decay luminosity would also need to
diffuse out through a massive envelope.

\begin{figure*}
\epsscale{0.84}
\plotone{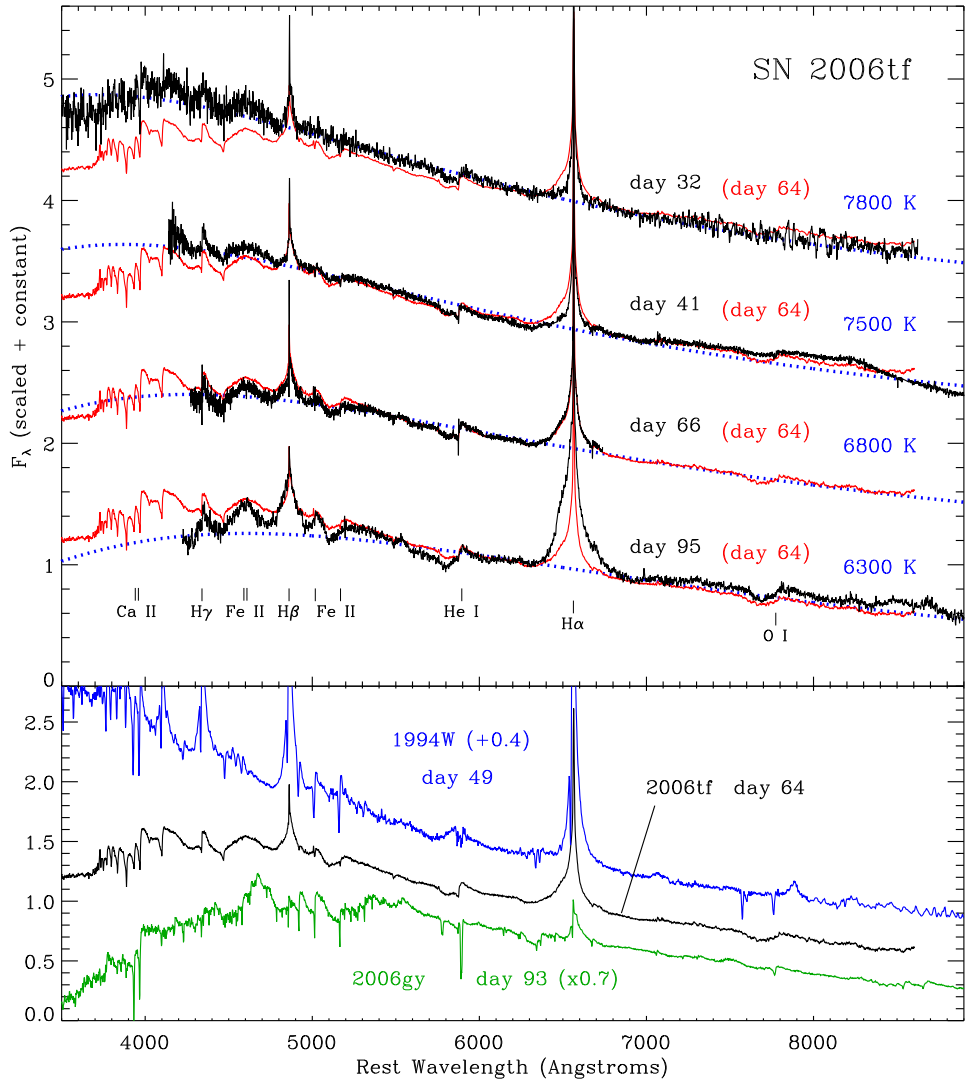}
\caption{The top panel shows visual-wavelength spectra of SN~2006tf on
days 32, 41, 66, and 95 since discovery, normalized to the red
continuum level, and with constant offsets of +3 (d32), +2 (d41), and
+1 (d66).  The day 64 spectrum (obtained for spectropolarimetry) is
plotted in red several times for comparison with the other epochs.
The dotted blue curves show representative black bodies for comparison
at the temperatures indicated.  At the bottom, the day 64 spectrum of
SN~2006tf (black) is compared to the day 93 spectrum of SN~2006gy
(green), dereddened assuming $E(B-V)=0.723$ mag (Smith et al.\ 2007),
and to the day 49 spectrum of SN~1994W (blue) from our spectral
database.  All spectra of SN~2006tf have been corrected for
$E(B-V) = 0.027$ mag.}\label{fig:allspec}
\end{figure*}

In this paper we investigate yet another extremely luminous SN that is
nearly as luminous as SNe 2006gy and 2005ap and has some similarity to 
them, but which exhibits signs of stronger CSM interaction in the spectrum.

SN~2006tf was discovered on 2006 Dec. 12 (UT dates are used
throughout this paper) in the course of the Texas Supernova Search
(Quimby et al.\ 2007b).  In spectra obtained $\sim$10~d after
discovery it showed Type IIn signatures.  SN~2006tf is located
0$\farcs$2 east and 0$\farcs$7 north of the center of an anonymous and
faint galaxy with an apparent $R$ magnitude of 20.68 (Quimby et al.\
2007b), for which the metallicity is not known.  Its redshift, $z =
0.074$, indicates a distance of 308 Mpc, assuming H$_0 = 72$ km
s$^{-1}$ Mpc$^{-1}$.  At that distance, the discovery magnitude of
about $m_r = 16.7$, which was roughly constant for the first 2 weeks
after discovery, makes the peak absolute magnitude about $-$20.7, or
roughly 1.6$\times$10$^{10}$~L$_{\odot}$ without correcting for
extinction.

This extremely high luminosity makes SN~2006tf one of the most
luminous SNe ever discovered.  Among known SNe, it is surpassed only
by SN~2006gy and SN~2005ap.  It is more luminous than very bright or
long-lasting SNe~IIn with strong CSM interaction like SNe~1979C,
1988Z, 1994W, 1998S, and others, as well as the Type Ia/IIn ``hybrid''
objects mentioned earlier.  Given its Type~IIn spectrum, the extreme
luminosity of SN~2006tf is likely to be powered by CSM interaction.
With the lack of any Type IIn signatures in the spectrum of SN~2005ap
(Quimby et al.\ 2007a), and the difficulty in reconciling the energy
budget of SN~2006gy with its relatively weak CSM interaction
signatures (Smith et al.\ 2007, 2008b), the strong CSM signatures and
comparably high luminosity of SN~2006tf are of significant interest.
Therefore, in this paper we consider the photometric and spectroscopic
evolution of SN~2006tf in detail.

We present our optical observations in \S 2. The
light curve and bolometric luminosity are discussed in \S 3, and in \S
4 we describe the general spectral evolution.  A detailed look at the
H$\alpha$ and H$\beta$ emission-line profiles and their evolution is given in
\S 5, including a late-time spectrum taken on day 445. Analysis of our
spectropolarimetry of SN~2006tf is presented in \S 6.  In \S 7 we
estimate the most basic physical properties of SN~2006tf. Section 8
summarizes our main results, discussed in the context of an integrated
picture of SN~2006tf and other SNe powered by CSM interaction.

\begin{deluxetable*}{lcccccccc}\tabletypesize{\scriptsize}
\tablecaption{KAIT Magnitudes of SN 2006\lowercase{tf}\tablenotemark{a}}
\tablewidth{0pc}
\tablehead{
  \colhead{MJD} &\colhead{$B$}  &\colhead{err$_B$} &\colhead{$V$} &\colhead{err$_V$} 
  &\colhead{$R$} &\colhead{err$_R$} &\colhead{$I$} &\colhead{err$_I$}
}
\startdata
2454115.02  &17.739  &0.030  &17.199  &0.017  &16.923  &0.014  &16.707  &0.028  \\   
2454118.99  &17.784  &0.036  &17.322  &0.028  &17.040  &0.021  &16.711  &0.052  \\   
2454121.99  &17.852  &0.041  &17.300  &0.030  &17.032  &0.030  &16.767  &0.039  \\   
2454125.00  &17.794  &0.025  &17.338  &0.019  &17.036  &0.014  &16.754  &0.031  \\   
2454131.00  &17.773  &0.039  &17.379  &0.031  &17.123  &0.023  &16.720  &0.040  \\   
2454133.94  &\nodata &\nodata &17.408 &0.079  &\nodata &\nodata &16.769 &0.129  \\   
2454144.04  &18.183  &0.075  &17.601  &0.153  &17.177  &0.145  &16.730  &0.107  \\   
2454146.97  &18.233  &0.075  &\nodata &\nodata &17.245 &0.072  &16.895  &0.035  \\   
2454149.94  &18.184  &0.033  &17.517  &0.023  &17.179  &0.018  &16.892  &0.039  \\   
2454161.98  &18.275  &0.370  &17.549  &0.080  &17.291  &0.054  &16.886  &0.070  \\   
2454170.92  &18.461  &0.045  &17.837  &0.032  &17.455  &0.027  &16.987  &0.055  \\   
2454175.90  &18.486  &0.047  &17.796  &0.031  &17.488  &0.028  &17.125  &0.045  \\   
2454178.88  &18.598  &0.049  &17.858  &0.047  &17.571  &0.025  &17.119  &0.044  \\   
2454181.86  &18.675  &0.058  &17.926  &0.028  &17.638  &0.031  &17.226  &0.037  \\   
2454184.88  &18.627  &0.083  &18.115  &0.127  &17.614  &0.068  &17.235  &0.100  \\   
2454188.86  &18.527  &0.092  &18.153  &0.064  &17.621  &0.037  &17.196  &0.059  \\   
2454198.84  &18.814  &0.168  &18.231  &0.126  &17.849  &0.152  &17.336  &0.116  \\   
2454201.80  &19.004  &0.086  &18.146  &0.045  &17.865  &0.029  &17.426  &0.041  \\   
2454204.85  &\nodata &\nodata &18.252 &0.148  &17.889  &0.257  &17.549  &0.136  \\   
2454207.82  &19.120  &0.073  &18.368  &0.062  &17.977  &0.043  &17.628  &0.058  \\   
2454211.81  &19.186  &0.088  &18.348  &0.053  &18.012  &0.030  &17.647  &0.055  \\   
2454225.81  &19.003  &0.097  &18.547  &0.064  &18.113  &0.046  &17.765  &0.087  \\   
2454230.77  &19.281  &0.061  &18.626  &0.055  &18.239  &0.041  &17.867  &0.051  \\   
2454235.79  &19.280  &0.120  &18.610  &0.122  &18.165  &0.080  &17.976  &0.071  \\   
2454242.75  &19.710  &0.071  &18.798  &0.075  &18.401  &0.062  &17.967  &0.121  \\   
2454249.71  &19.475  &0.137  &18.767  &0.137  &18.589  &0.160  &18.004  &0.192  \\   
2454254.77  &19.442  &0.248  &18.767  &0.127  &18.374  &0.101  &\nodata &\nodata \\   
\enddata
\tablenotetext{a}{Photometric uncertainties are 1$\sigma$.}
\end{deluxetable*}

\begin{deluxetable*}{lclcccccc}\tabletypesize{\scriptsize}
\tablecaption{Spectroscopic Observations of SN~2006\lowercase{tf}}
\tablewidth{0pt}
\tablehead{
  \colhead{Date} &\colhead{Day\tablenotemark{a}} &\colhead{Instrument}
  &\colhead{Range\tablenotemark{b}} &\colhead{$\lambda$/$\Delta\lambda$} 
  &\colhead{EW(H$\alpha$)}  &\colhead{$F$(H$\alpha$)} 
  &\colhead{H$\alpha$/H$\beta$\tablenotemark{c}} &\colhead{N/T\tablenotemark{d}} \\
  \colhead{} &\colhead{} &\colhead{} &\colhead{\AA} &\colhead{} &\colhead{\AA}
  &\colhead{10$^{-14}$ erg s$^{-1}$ cm$^{-2}$} &\colhead{} &\colhead{\%} 
}
\startdata
2007~Jan.~13  &32  &LRIS/Keck~I    &3500--8600  &1000 &$-$44(6)   &1.33(0.1)  &2.85  &17   \\
2007~Jan.~22  &41  &DEIMOS/Keck~II &4200--8900  &2000 &$-$66(5)   &1.81(0.17) &3.67  &11.2 \\
2007~Feb.~14  &64  &LRIS/Keck~I~(pol) &3500--8600 &700 &$-$72(5)  &2.19(0.16) &3.48  &9.0  \\
2007~Feb.~16  &66  &DEIMOS/Keck~II &4300--6750  &8000 &$-$82(7)   &1.90(0.2)  &3.79  &8.1  \\
2007~Mar.~17  &95  &DEIMOS/Keck~II &4200--8900  &2000 &$-$194(11) &3.39(0.24) &4.84  &3.4  \\
2008~Mar.~1 &445 &ESI/Keck~II &4400--8800 &6000 &$-$1035(90)      &0.69(0.34) &11.6 &3.6 \\
\enddata
\tablenotetext{a}{Days since discovery.  The additional number of days 
since explosion could be as much as 50--70~d for SN~2006tf, if the
behavior of SN~2006gy is relevant.}
\tablenotetext{b}{Spectral range in rest wavelength.}
\tablenotetext{c}{The H$\alpha$/H$\beta$ flux ratio measured after the
spectra are dereddened.}
\tablenotetext{d}{Percentage contribution of the narrow component to
the total H$\alpha$ line flux.}
\end{deluxetable*}

\section{OBSERVATIONS}

\newcommand{\ubvri}{\protect\hbox{$U\!BV\!RI$} }
\newcommand{\bvri}{\protect\hbox{$BV\!RI$} }

We obtained optical photometry of SN 2006tf in \bvri using the 0.76~m
Katzman Automatic Imaging Telescope (KAIT; Filippenko et al. 2001) at
Lick Observatory. Flat fielding and bias subtraction were processed
automatically. Galaxy subtraction and differential photometry were
done using the KAIT pipeline (Ganeshalingam et al., in
prep.). Template images of the galaxy were obtained on 2008~Jan.~18
with KAIT, more than 1 yr after discovery.  Our late-time spectrum
(see below) shows strong H$\alpha$ emission and some faint continuum
detected from the SN.  Hence, there is some concern that our template
image obtained on day 404 is partly contaminated by SN emission,
rather than pure background galaxy emission.  At this late time, we
measure the magnitude at the position of the SN to be 20.1 in the $R$
band.  Even if this late-time magnitude is dominated by SN light, the
consequent oversubtraction would have caused us to overestimate the
true $R$ magnitude by only 0.22 mag after day 150, which is not much
larger than the photometric uncertainty at that time.

Calibrations for the field were obtained on 8 photometric nights using
both KAIT and the 1~m Nickel telescope at Lick Observatory. 
The uncertainty in our subtraction and photometry pipeline is estimated by
injecting artificial stars with the same magnitude and point-spread function
of the SN into regions of comparable complexity in the original KAIT images and
recovering them. The final uncertainty is taken to be the scatter in
recovering 20 artificial stars added in quadrature with the
calibration error.

We take day zero to be MJD = 2,454,081.98, the discovery date 2006~Dec.~12
(Quimby et al.\ 2007b).  Since this is not the
explosion date, there will be a source of continual ambiguity when
comparing SN~2006tf to other SNe.  The explosion date is not known,
but if the light-curve evolution was similar to that of SN~2006gy, the
explosion of SN~2006tf could conceivably have been long before the
time of discovery.  Figure~\ref{fig:mags} shows the KAIT magnitudes,
which are listed in Table~1.

During the main peak of the light curve (within 200~d of discovery),
we obtained visual-wavelength spectra of SN~2006tf on five separate
dates at the Keck Observatory using the low resolution imaging
spectrometer (LRIS; Oke et al.\ 1995) and the Deep Imaging
Multi-Object Spectrograph (DEIMOS; Faber et al.\ 2003), with the
observations summarized in Table 2.  The epochs for these four spectra
are days 32, 41, 64, 66, and 95 after discovery.  The spectra were all
obtained with the long slit oriented at the parallactic angle
(Filippenko 1982), and were reduced using standard techniques.
Wavelengths were corrected for redshift $z = 0.074$ so that the narrow
H$\alpha$ emission line was at the proper rest wavelength.  Given the
very weak Na~{\sc i}~D absorption and blue color, we made no
correction for reddening and extinction beyond the relatively small
Galactic values of $E(B-V) = 0.027$ mag and $A_R = 0.062$ mag
(Schlegel et al.\ 1998).  The resulting spectra are shown in
Figure~\ref{fig:allspec}.

We also obtained one late-time optical spectrum on 2008~March~1, day
445 after discovery, using the Echelle Spectrograph and Imager (ESI;
Sheinis et al. 2002) on the Keck~II telescope (see Table 2).  A 1200~s
exposure was obtained in echellete mode, with the 0$\farcs$75-wide
slit oriented at position angle 45$\arcdeg$.  Because of relatively
poor seeing of $\sim$1$\farcs$5 (wider than the slit) and possible
light cirrus during the night, the flux calibration is uncertain, and
we conservatively adopt a factor of 2 uncertainty in absolute flux
(i.e., H$\alpha$ in Table 2).  This is larger than the
line-measurement uncertainty, and is hard to quantify.  The late-time
spectrum is discussed in \S 5.4.

Thus, we have two independent estimates of the late-time $R$
magnitude.  As we mentioned earlier, our KAIT template image shows a
magnitude of 20.1, or $L = 7.5 \times 10^8$ L$_{\odot}$, at the
position of the SN.  The true SN light could be substantially fainter
than this if background galaxy light is important, but it cannot be
much brighter (note that H$\alpha$ also contaminates this
measurement).  From our late-time spectrum on day 445, in which we
subtracted nearby background and still detected faint continuum
emission from the SN, we measure $F_{\lambda} = 1.4$ ergs s$^{-1}$
cm$^{-2}$ \AA$^{-1}$ for the continuum level at red wavelengths, or an
apparent $R$ magnitude of 20.3. This corresponds to a late-time
bolometric luminosity of $L = 6.3 \times 10^8$ L$_{\odot}$.  Even
though our uncertainty in the flux is a factor of two because of sky
conditions, the agreement with the late-time KAIT magnitude to better
precision than that is reassuring.

Of the five epochs of spectra during the first 100~d, we obtained
one epoch of spectropolarimetric observations of SN~2006tf on
2007~Feb.~14.54 (day 64) using the polarimeter unit\footnote{See
\url{http://alamoana.keck.hawaii.edu/inst/lris/polarimeter/} for the
online polarimeter manual by Cohen (2005).} of LRIS.  The object was
observed for 1000~s in each of the four rotation angles of the
half-wave plate retarder.  The total observed spectral range was
3350--9240~\AA, and the 1$\farcs$5-wide slit gave a spectral
resolution of $\sim$5~\AA \ in the blue and $\sim$9~\AA \ in the red.
The spectropolarimetric reductions followed the procedure outlined by
Miller et al.\ (1988) and implemented by Leonard et al.\ (2001).
Observations of the polarized standard stars HD~19820 and HD~161056
(Schmidt et al.\ 1992) gave consistent results for the zero point of
the instrument.  Instrumental polarization was negligible as both of
the observed null polarization standards showed $<$0.1\% polarization.
Two foreground stars (BD~+11~2491, BD~+12~2510), both within
20\arcmin\ of the SN, were observed to constrain any potential
Galactic component of interstellar polarization (ISP) due to dust
grains along the line of sight.  Both stars showed less than 0.1\%
polarization, indicating that the Galactic component of ISP is also
negligible, consistent with the low dust column of $E(B-V) = 0.027$
mag at the high Galactic latitude of SN~2006tf ($b =$ 74\degr).

\begin{figure}
\epsscale{1.1} 
\plotone{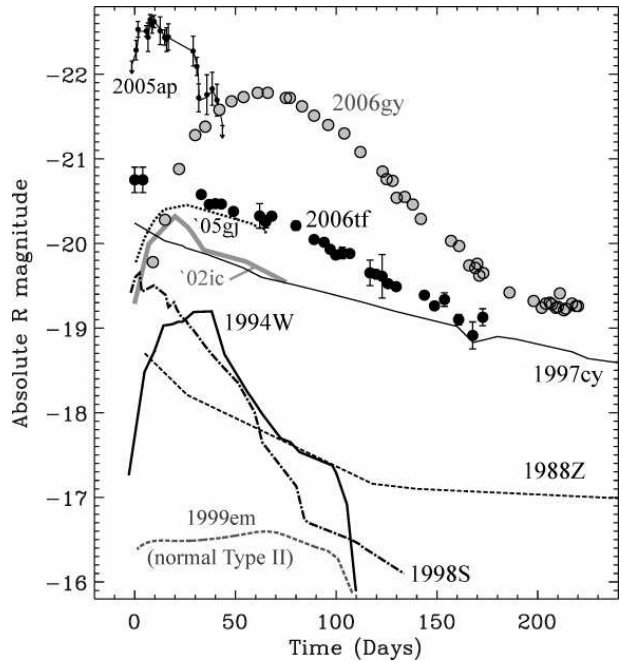}
\caption{The absolute $R$-band light curve of SN~2006tf compared to
that of several other luminous or strongly interacting SNe~II and IIn.
SN~2005ap is a Type II (not a IIn) from Quimby et al.\ (2007a) and
SN~1999em is a normal Type II-P SN (Leonard et al.\ 2002).  All others
shown are Type IIn: SN~2006gy data are from Smith et al.\ (2007),
SN~1994W data from Sollerman et al.\ (1998), SN~1988Z from Turatto et
al.\ (1993), and SN~1998S from Fassia et al.\ (2000).  SN~1997cy
(Germany et al.\ 2000) is an extremely luminous SN~IIn possibly
associated with GRB~970514, or alternatively, a possible ``hybrid''
Type Ia/IIn like SNe~2002ic (Wood-Vasey et al.\ 2004; Hamuy et al.\
2003) and 2005gj (Prieto et al.\ 2007) that are still difficult to
understand (see also Benetti et al.\ 2006).  When different values of
H$_0$ were used by the above authors, the absolute magnitudes have
been adjusted to correspond to our adopted value of 72 km s$^{-1}$
Mpc$^{-1}$.}
\label{fig:lightcurve}
\end{figure}

\section{THE LIGHT CURVE AND BOLOMETRIC LUMINOSITY}

Our \bvri photometry of SN~2006tf obtained with KAIT is presented in
Figure~\ref{fig:mags}.  Two features are immediately apparent.  The
first is the remarkably slow decline in the light curve, with roughly
0.01 mag d$^{-1}$ in the $R$ band, reminiscent of the slow evolution
in SN~2006gy as well as the very slow decline in strongly interacting
SNe~IIn like SNe~1988Z, 1997cy, and 2005gj.  The decline rate also
matches the expected decay rate of $^{56}$Co, and this is discussed in
more detail below.  The second feature is the gradual change in color.
The lower panel in Figure~\ref{fig:mags} shows the observed $B-V$
color, which gets redder at a rate of $\sim 0.0025 \pm 0.0003$
mag~d$^{-1}$. This slow evolution to redder colors is in agreement
with our spectra in Figure~\ref{fig:allspec}, in which the continuum
slope gradually and monotonically evolves to lower temperatures.  The
relatively weak color changes, with temperatures that imply little or
no bolometric correction, mean that we can take the absolute $R$
magnitude (corrected for small Galactic extinction) as a proxy for the
bolometric magnitude, to compare SN~2006tf with other luminous SNe.

Thus, in Figure~\ref{fig:lightcurve} we compare the absolute $R$ mag
of SN~2006tf to that of several other notable SNe~II/IIn.  Near its
peak at the time of discovery, SN~2006tf was the third most luminous
SN observed so far after SN~2005ap and SN~2006gy.  However, SNe
2005gj, 2002ic, and 1997cy are not far behind.  All of these are
SNe~II.  Interestingly, though, all {\it except} the most luminous,
SN~2005ap, were SNe~IIn with prominent and relatively narrow H$\alpha$
indicating strong CSM interaction.\footnote{The true natures of
SNe~2002ic, 2005gj, and 1997cy are still controversial: SNe~Ia
exploding in dense H envelopes or perhaps the deaths of massive stars;
see Benetti et al.\ (2006).}  The strength of the SN~IIn signature may
be an important clue to the nature of the most luminous SNe: As we
noted in a previous paper (Smith et al.\ 2007), while SN~2006gy was
definitely a Type~IIn with an H$\alpha$ profile that indicated {\it
some} CSM interaction, the {\it strength} of that interaction
indicated by the H$\alpha$ and X-ray luminosity seemed far too weak to
be compatible with the power in continuum light.  SN~2005ap was an
even more extreme case, with no detectable sign of a narrow H$\alpha$
feature generated by CSM interaction.  SN~2006tf may be an important
link, as its continuum luminosity and strong H$\alpha$ are
intermediate between these two extreme cases and other strongly
interacting SNe.  This will be a recurring theme in our investigation.

SN~2006tf was apparently discovered at, or perhaps somewhat after, its
time of peak luminosity.  The explosion date and the photometric
behavior during the rise to maximum are not known. This should be kept
in mind when interpreting Figure~\ref{fig:lightcurve}, because the
translation of SN~2006tf on the time axis is necessarily quite
uncertain.  Judging by its similarity to SN~2006gy, for example,
SN~2006tf could potentially be shifted as much as 50--70~d to the
right to be more correctly compared with other SNe plotted here.
Indeed, its high luminosity requires a large emitting radius of order
300 AU.  That, combined with its maximum observed expansion speeds of
order 7500 km s$^{-1}$ seen in the broad component of the H$\alpha$
line (see below), argue that it may have had quite a long rise time of
order 50--100~d.  Judging by SN~2002ic and SN~2005gj, both of which
were caught during the rise to maximum, SN~2006tf should be shifted to
the right in Figure~\ref{fig:lightcurve} by {\it at least} 20~d.

In any case, SN~2006tf clearly has an extremely high peak luminosity
and a very long duration, with no sign, as yet, of a sharp decline
like that of SNe~1994W or 2005ap (Fig.~\ref{fig:lightcurve}).  As
such, it demands an explanation for its high fluence that far exceeds
that of normal core-collapse SNe.  In general, such longevity at high
luminosity implicates a large mass of emitting material, as we shall
see below.  Although less extreme, the energy demands of SN~2006tf
rival those of SN~2006gy, and considerations for the possible energy
source are similar.  More in-depth considerations of the possible
energy sources and associated difficulties can be found in our
previous papers on SN~2006gy (Smith et al.\ 2007, 2008b; Smith \&
McCray 2007).  The two sources of luminosity we consider here are
$^{56}$Co decay from a large initial mass of $^{56}$Ni, or
shock-deposited kinetic energy diffusing out of an opaque envelope
after being thermalized and converted to visual light.

\begin{figure}
\epsscale{1.0} 
\plotone{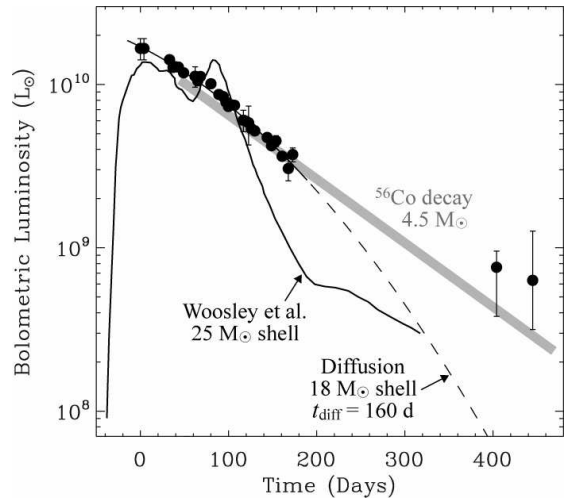}
\caption{The absolute $R$ magnitudes of SN~2006tf from
Figure~\ref{fig:lightcurve} are converted to luminosities in
L$_{\odot}$; we also include the uncertain late-time magnitudes
mentioned in \S 2.  These are compared to a simple photon diffusion
model (Smith \& McCray 2007) and the decline rate of $^{56}$Co decay.
The diffusion model (long dashes) gives an acceptable fit to the
light curve for a diffusion timescale $t_{\rm diff} = 160$~d and a CSM
shell of $\sim$18 M$_{\odot}$, while radioactive decay (thick gray
line) would require $\sim$4.5 M$_{\odot}$ of $^{56}$Ni to power the
same luminosity (see text).  We also show a CSM interaction model for
a progenitor star of 110 M$_{\odot}$ and a circumstellar envelope mass
of $\sim$25 M$_{\odot}$ (Woosley et al.\ 2007).}\label{fig:diffusion}
\end{figure}

\subsection{Radioactive Decay?}

One possible source for a long lasting, extremely high SN luminosity
is radioactive decay from a large initial mass of $^{56}$Ni.  This
might be synthesized in a ``hypernova'' or collapsar as has been
proposed in the case of SN~1997cy where 2.3--2.6 M$_{\odot}$ of
$^{56}$Ni was derived (Germany et al.\ 2000; Turatto et al.\ 2000); in
that case, it is interesting that the light curve of SN~1997cy is
quite similar to that of SN~2006tf.  Alternatively, a large mass of
$^{56}$Ni could potentially be synthesized in a pair-instability SN
explosion (Bond et al.\ 1984; Barkat et al.\ 1967; Rakavy \& Shaviv
1967), as had been noted as a possibility for SN~2006gy (Smith et al.\
2007) and SN~2005ap (Quimby et al.\ 2007a).

Figure~\ref{fig:diffusion} shows the bolometric luminosity light curve
of SN~2006tf, derived from the absolute $R$ magnitude (adopting
$A_R = 0.062$ mag).  The thick gray line shows the $^{56}$Co luminosity
and decay rate one would expect from an initial $^{56}$Ni mass of
about 4.5 M$_{\odot}$.\footnote{The necessary $^{56}$Ni mass is very
uncertain. It could be 7 M$_{\odot}$ if our adopted $t = 0$ is actually
$\sim$50~d after the explosion date, or it could be arbitrarily
less if CSM interaction contributes some substantial fraction of the
luminosity.}  Although it does not match the light curve precisely,
radioactive decay does reproduce the average decline rate of SN~2006tf
quite well.  It gives as good an account of the light curve as the
2.3--2.6~M$_{\odot}$ of $^{56}$Ni proposed in the case of SN~1997cy
(Germany et al.\ 2000; Turatto et al.\ 2000).

Furthermore, radioactive decay makes a clear prediction for the
late-time luminosity: it should follow the $^{56}$Co decay rate.  As
we noted earlier, on day 404, a source at the position of SN~2006tf
had an $R$ magnitude of roughly 20.1 (or somewhat fainter) in the
template image that we used to subtract background light and calibrate
our KAIT photometry in Figure~\ref{fig:mags}. Also, from the continuum
level in a late-time spectrum obtained on day 445, we estimate an
apparent $R$ magnitude of $20.3 \pm 0.75$, as noted above. The late-time
luminosity predicted by the same $^{56}$Co decay from the earlier
light curve is roughly consistent with these late-time measurements.
The decay model actually underpredicts the luminosity, but so do the
CSM interaction models in Figure~\ref{fig:diffusion}, and hence {\it any}
model requires us to invoke some additional CSM interaction at late
times.

Thus, we cannot easily discount $^{56}$Co decay as a potentially
significant source of luminosity for SN~2006tf, especially if some
portion of the luminosity is supplied by CSM interaction as indicated
by the spectral characteristics.  This was also the case for SN~1997cy
and SN~2006gy.  Although we cannot confidently rule out radioactive
decay, it seems more likely to us that SN~2006tf is dominated by
extreme CSM interaction.

\subsection{Photon Diffusion From a Massive Shocked Envelope?}

A better match to the main light-curve peak of SN~2006tf is attained
with an opaque shell-shocked model like that which Smith \& McCray
(2007) proposed for the main light-curve peak of SN~2006gy
(Fig.~\ref{fig:diffusion}). In such a model, the extreme luminosity is a
product of the large initial radius of an optically thick CSM ejecta
envelope, throughout which shock kinetic energy has been thermalized
efficiently.  If the envelope is highly opaque, radiation must diffuse
out rather than being emitted directly by the post-shock cooling zone.
{\it This is true no matter what the ultimate source of energy
deposition is}, whether it be shock energy or radioactive decay, or
both.

The main ingredients that determine the shape of the diffusion light
curve are the characteristic time for photon diffusion, $t_{\rm diff}
= \tau R/c$, and the expansion time, $t_{\rm exp}$ (see Smith \&
McCray 2007; Falk \& Arnett 1973, 1977).  These can be expressed as
$t_{\rm diff} \approx 23 (M/R_{15})$~d, with $M$ being the envelope
mass in M$_{\odot}$ and $R_{15}$ being the characteristic initial
shell radius in units of 10$^{15}$ cm.  The expansion timescale is
just $t_{\rm exp} \approx (\Delta R/v_{\rm exp})$ for constant
expansion speed.  Our model (dashed curve in Fig.~\ref{fig:diffusion})
is a simple analytic model with $t_{\rm diff} = 160$~d, $R_{15} = 2.7$,
and $M$=18~M$_{\odot}$.  For comparison, Figure~\ref{fig:diffusion}
also shows a one-dimensional (1-D) numerical model of a similar
physical situation from Woosley et al.\ (2007), with a 25 M$_{\odot}$
shell ejected by a 110 M$_{\odot}$ progenitor star about 5 yr before
the SN.  This model was intended for SN~2006gy, but seems even more
applicable to SN~2006tf.\footnote{Woosley et al.\ (2007) note that the
peaks in the light curve may be an artifact of their 1-D models, which
they expect would be smoothed out in reality.}

The diffusion curve in Figure~\ref{fig:diffusion} provides an
excellent fit to the observed data by adopting $t_{\rm diff} = t_{\rm
exp} = 160$~d (maximum luminosity for a given mass is achieved when
these two are equal, minimizing adiabatic losses).  If the
characteristic expansion speed of the blast wave is about 2,000 km
s$^{-1}$, as indicated by the intermediate-width H$\alpha$ emission
(see below), then $R_{15} = 2.7$ or about 180 AU for $t_{\rm exp} =
160$~d.  This is the initial radius of the shocked ejecta envelope,
which will grow with time.  Thus, it is in approximate agreement with
the somewhat larger radius of $\sim$300 AU that we would nominally
expect for a $\sim$7800~K black body emitting the peak luminosity of
SN~2006tf (see below).  For $t_{\rm diff} = 160$~d, the required
envelope mass is $\sim$18 M$_{\odot}$.  These values are similar to,
but even more extreme than, those of SN~2006gy: $t_{\rm diff} = 70$~d, $R
= 160$~AU, and $M = 10$~M$_{\odot}$, respectively (Smith \& McCray
2007).  As in that case, this radius is far too large to be the actual
hydrostatic radius of a red supergiant progenitor, and the CSM-wind
expansion speed of 190 km s$^{-1}$ noted earlier is much too fast for
that interpretation as well.  Instead, this large radius probably
represents the pseudo-photosphere of an opaque pre-SN ejecta shell
that is not bound to the star, having been ejected $\sim$4 yr prior to
the SN if it has been moving at $\sim$190 km s$^{-1}$.  Containing
18~M$_{\odot}$ within a radius of 180 AU would give at least
$\tau \approx 400$.

In this model, a self-consistent explanation for the lower peak
luminosity and slower decline rate of SN~2006tf as compared to
SN~2006gy is that the forward shock of SN~2006tf has a slower expansion
speed of 2,000 km s$^{-1}$, compared to 4,000 km s$^{-1}$ for SN~2006gy
(Smith et al.\ 2007).  The slower expansion speed causes a lower peak
luminosity and longer duration, because the same amount of thermal
energy will take a longer time to leak out of the envelope and the
radius is smaller at a given time after explosion.

Although a simple model like this can fit the main light-curve peak,
it is not necessarily a unique explanation, and both types of CSM
interaction models (Smith \& McCray 2007; Woosley et al.\ 2007) fall
short of the observed late-time luminosity (Fig.~\ref{fig:diffusion}).
As with SN~2006gy, there must be some additional contribution of
emission directly from the ongoing, more optically thin CSM
interaction region in order to explain the strong H$\alpha$ emission.
This occurs subsequent to the shock wave passing through the massive
opaque shell, before which time the post-shock H$\alpha$ emission
would not be seen.

\subsection{Total Radiated Energy}

Regardless of which interpretation is correct, the photon diffusion
model in Figure~\ref{fig:diffusion} gives an accurate phenomenological
fit to the bolometric luminosity, so it can be used to measure the
total radiated energy.  Integrating this curve in
Figure~\ref{fig:diffusion} from day 0 to 180, we find that the total
energy radiated in visual light during the time of our photometric
monitoring was ($6.2 \pm 0.3) \times 10^{50}$ ergs, if we assume no
bolometric correction to the absolute $R$ magnitude.  If we include
the late-time tail, this value rises to about $7.0 \times 10^{50}$
ergs, and it could be increased even further if the time of discovery
occurred significantly {\it after} the actual time of explosion.  This
is likely to be the case if SN~2006tf had a slow rise time comparable
to SN~2006gy, or even a faster rise akin to SNe~2002ic and 2005gj, as
we noted earlier.

It is interesting to note that an 18~M$_{\odot}$ shell moving at 2000
km s$^{-1}$ contains the same amount of kinetic energy of about $7
\times 10^{50}$ ergs.  Rough equipartition of thermal and kinetic
energy is not surprising in this model (see Smith \& McCray 2007).

In any case, the total amount of energy radiated in visual light is
almost 10$^{51}$ ergs, or about half that of SN~2006gy (Smith et al.\
2007).  This amount of radiated energy must drain the reservoir of
total available kinetic energy.  Since the SN showed no sign of
deceleration, this suggests either (1) that the CSM interaction we see
now is not the major power source for the light curve, occurring
subsequent to the shock passing through the opaque envelope mentioned
earlier, or (2) that SN~2006tf marked an unusually energetic explosion
well in excess of the canonical kinetic energy of a SN, as was the
case with SN~2006gy (see Smith et al.\ 2007 for further discussion).

\section{THE VISUAL-WAVELENGTH SPECTRUM AND CONTINUUM}

The top panel in Figure~\ref{fig:allspec} shows our visual-wavelength
spectra of SN~2006tf on the four dates during the decline from the
main peak of the light curve, normalized to the level of the red
continuum flux and offset by constant values as noted in the caption.
The epoch on day 64 is repeated in red for comparison to the other 4
epochs (this epoch has the highest signal-to-noise ratio because it was
obtained for spectropolarimetry; see \S 6).  These spectra have been
corrected for Galactic reddening of $E(B-V) = 0.027$ mag.

One of the most striking changes with time is in the relative strength
of H$\alpha$, to be discussed in detail in the following
section.  Here we focus mainly on the continuum shape and overall
properties of the spectra.

In general, Figure~\ref{fig:allspec} reinforces our earlier conclusion
from photometry that SN~2006tf shows mild change in its color and
continuum shape during the first 100~d after discovery, with
correspondingly little evolution in the character of the spectrum.
Matching the continuum with black bodies (dotted blue curves in the top
panel of Fig.~\ref{fig:allspec}), the characteristic temperature drops
monotonically from 7800~K to 6300~K between days 32 and 95, consistent
with the steady change in $B-V$ (Fig.~\ref{fig:mags}).

\begin{figure}[ht]
\epsscale{1.2} 
\plotone{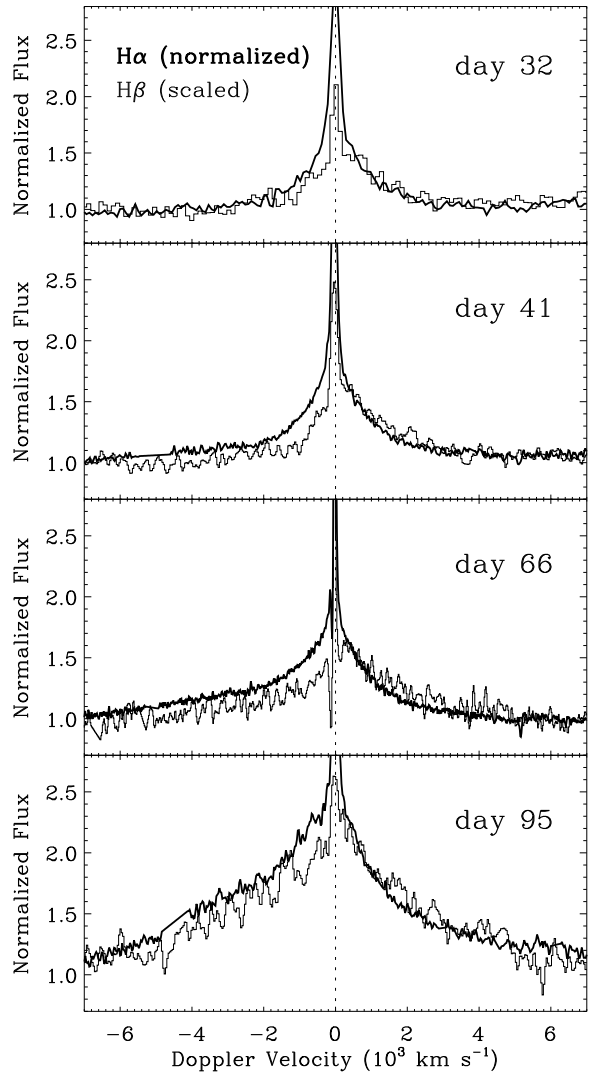}
\caption{Observed line profiles of H$\alpha$ and H$\beta$ in
  SN~2006tf.  H$\alpha$ has been normalized with a fit to the observed
  continuum level.  H$\beta$ for each date is normalized and then
  scaled arbitrarily for comparison with the H$\alpha$ profile.  Day
  64 is not shown here or in the next few figures because it is only 2~d 
  before day 66 and has lower spectral resolution.}
\label{fig:hahb}
\end{figure}

Interestingly, the slow drop in temperature and characteristic
line-blanketing features shortward of 4500 \AA\ are atypical for
strongly interacting SNe~IIn, which tend to show very blue continua at
higher temperatures that remain nearly constant if the blast-wave
speed is constant.  Instead, the slow decline in temperatures around
7000~K and pronounced line blanketing are more reminiscent of normal
SNe II-P, where photons are diffusing out of the expanding SN ejecta
through which the recombination photosphere is receding.  Since
SN~2006tf is far too luminous for this emission to arise in normal SN
ejecta, this is another argument favoring the shell-shocked photon
diffusion model that we mentioned earlier, and the sustained high
luminosity means that this shell must be very massive.  SN~2006tf is
distinguished from SNe~II-P, however, in that its observed constant
expansion speed indicates that this process must occur in a thin
expanding shell rather than the geometrically thick envelope with a
large velocity gradient in SNe~II-P.

In the bottom panel of Figure~\ref{fig:allspec} we compare the
spectrum of SN~2006tf (black) to that of two other well-studied
SNe~IIn from our spectral database.  SN~1994W (blue) is a case where
the spectrum and luminosity are dominated by ongoing CSM interaction
with a contribution from diffusion of radiation from the SN ejecta or
a shocked opaque envelope (Chugai et al.\ 2004), whereas in SN~2006gy
(green), most of the continuum luminosity arises from photon diffusion
from a previously shocked opaque shell (Smith \& McCray
2007).\footnote{The fact that we are comparing a day 64 spectrum of
SN~2006tf to a day 93 spectrum of SN~2006gy is not necessarily cause
for concern, since one is the date since discovery, and the other is
the much longer time since the presumed explosion date.  They
represent comparable times after peak luminosity.}  The continuum
shape of SN~2006gy matches that of SN~2006tf quite well, as noted by
Smith et al.\ (2007), but the H$\alpha$ line of SN~2006tf is much
stronger relative to its continuum.  Taking into account both the
continuum shape and the strong Balmer-line emission, we find that a
linear combination of the spectra of both SN~1994W and SN~2006gy gives
a suitable approximation to SN~2006tf.  This suggests that ongoing CSM
interaction, as compared to diffusion, contributes a larger fraction
of the observed luminosity in SN~2006tf than in SN~2006gy.
Furthermore, the increase in the relative strength of the broad
component of H$\alpha$ with time implies that CSM interaction
contributes a larger fraction of the luminosity at late times as the
continuum source fades.  This makes a strong case that SN~2006tf is a
valuable example of the transition region between opaque shocked
shells, like in SNe~2006gy and 2005ap, and progenitors with lower
mass-loss rates where CSM-interaction is seen directly, like in
SN~1988Z.

\begin{figure}[ht]
\epsscale{1.0}
\plotone{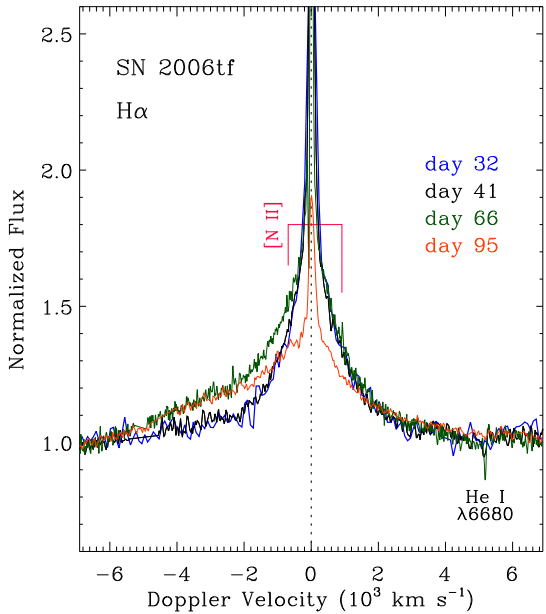}
\caption{Observed H$\alpha$ profiles on the four dates indicated,
  normalized and superposed over one another.  The first three
  dates are plotted without further scaling, but the last one on day
  95 has been adjusted by dividing the normalized intensity by 3 and
  adding 0.66 to compare the line shape with the other dates.
  Velocities at which [N~{\sc ii}] $\lambda\lambda$6548, 6583 would be
  observed are indicated.}
\label{fig:hanorm}
\end{figure}

Finally, we note that aside from its H$\alpha$ emission, the spectrum
of SN~2006tf does not resemble the ``hybrid'' SNe~IIn/Ia objects such as
SNe~2002ic and 2005gj at any epoch, even though their light-curve
shapes and absolute magnitudes are similar at early times.  We detect
no evidence for features from an underlying SN~Ia spectrum.
Additionally, the required envelope mass of 18~M$_{\odot}$ and total
radiated energy of almost 10$^{51}$ ergs make a SN~Ia-powered
interpretation unlikely for SN~2006tf.

\section{EVOLUTION OF H$\alpha$ AND H$\beta$ EMISSION}

Figure~\ref{fig:hahb} shows both H$\alpha$ and H$\beta$ at all four
epochs, with the H$\alpha$ flux normalized to the nearby continuum,
and the H$\beta$ profile scaled for comparison.  It is evident that
H$\beta$ is more asymmetric than H$\alpha$, showing a persistent
deficit in the blueshifted emission wing at all epochs after day 32,
due to high optical depths and the stronger self-absorption in
H$\beta$.

The intermediate-width component of H$\alpha$ at roughly $\pm$2000 km
s$^{-1}$ is discussed in \S 5.1, while broad emission at relatively
high speeds is discussed in \S 5.2. In \S 5.3 we describe the profiles
of the narrowest component from the unshocked CSM, in \S 5.4 we
present the late-time spectrum, and in \S 5.5 we discuss the behavior
of the line luminosity with time.

\begin{figure}[ht]
\epsscale{1.0}
\plotone{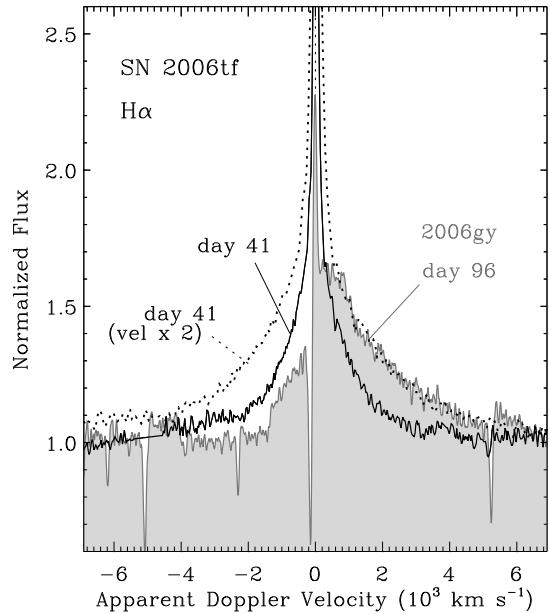}
\caption{Same as Figure~\ref{fig:hanorm} but comparing the day 41
spectrum of SN~2006tf (solid black) to the day 96 H$\alpha$ profile of
SN~2006gy (shaded gray, from Smith et al.\ 2007; the flux has been
scaled upward because the H$\alpha$ emission of SN~2006gy is much 
weaker relative to its continuum level).  The profile of SN~2006gy is
noticeably broader than that of SN 2006tf.  If we expand the velocity
scale of SN 2006tf by a factor of 2 (dashed line), then the red side of
the broad H$\alpha$ profile matches that of SN~2006gy quite well.
This suggests that the blast-wave speed of SN~2006tf is roughly half
as fast as that of SN~2006gy.}
\label{fig:ha06gy}
\end{figure}

\subsection{Post-Shock H$\alpha$ and H$\beta$ Profiles}

Figure~\ref{fig:hanorm} shows the intermediate-width H$\alpha$ profile
on days 32, 41, 66, and 95.  This component at $\pm$2000 km s$^{-1}$
dominates the appearance of the spectrum and accounts for most of the 
flux in the H$\alpha$ line.  The H$\alpha$ profile appears to be identical 
on days 32 and 41, and the red wing of the line maintains the same shape 
at all epochs.  However, we see a stronger, broad, blueshifted wing in
the day 66 and 95 spectra.  This enhanced blueshifted emission wing
makes the line appear asymmetric.  However, we suspect that it is not
related directly to the emission from post-shock gas, but instead is
due to an underlying broad component that is seen as the shell becomes
more effectively optically thin, as discussed in the next section.
Ignoring this broad blueshifted wing for now, the intermediate-width
component of H$\alpha$ is symmetric and centrally peaked in SN~2006tf,
typical of the H$\alpha$ profiles that define SNe~IIn.

The speed of the blast wave expanding into the CSM can be inferred
most clearly from the width and shape of this intermediate-width
H$\alpha$ profile.  Figure~\ref{fig:ha06gy} shows the H$\alpha$
profile of SN~2006tf on day 41, when it is most symmetric, and
compares it to the observed broad H$\alpha$ profile of SN~2006gy from
Smith et al.\ (2007).  The reason this is useful is because the broad
P~Cygni absorption feature in SN~2006gy has a sharp blue edge at
$-$4,000 km s$^{-1}$ that provides a good measure of the likely speed
of the blast wave (Smith et al.\ 2007), whereas the profile of
SN~2006tf has no sharp blue edge to provide a convenient measure of
its true speed.

\begin{figure}
\epsscale{0.95}
\plotone{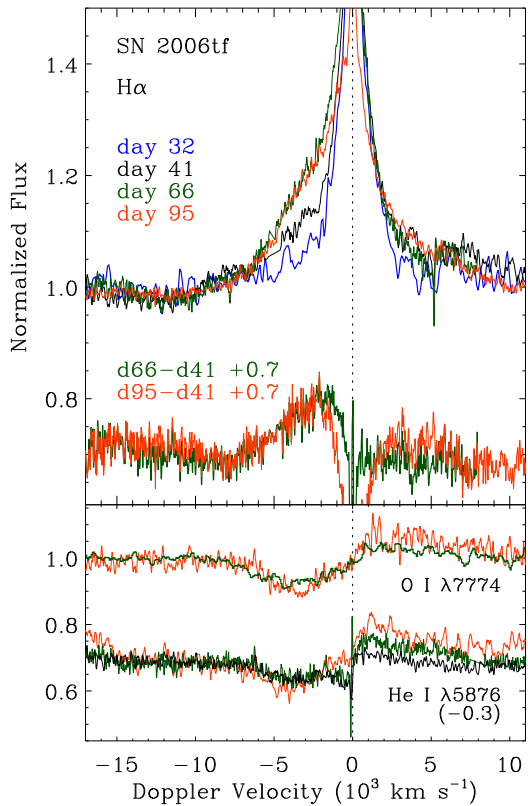}
\caption{({\it Top panel}) Same as Figure~\ref{fig:hanorm}, but with
an expanded velocity scale and smaller vertical range to show the
behavior of the broad blueshifted wings of H$\alpha$.  The lower
tracings show the day 66 and 95 residual profiles after subtracting
off the symmetric day 41 profile.  Residual fast blueshifted emission
is seen out to roughly $-$7500 km s$^{-1}$.  ({\it Bottom panel}) Line
profiles of O~{\sc i} $\lambda$7774 on day 64 (green) and 95 (orange),
and the feature at $\sim$5900 \AA\ (colors for different epochs are
the same as the top panel), which could potentially be either He~{\sc
i} $\lambda$5876 or Na~{\sc i} $\lambda$5892.  It is plotted here as
if it were He~{\sc i} $\lambda$5876 (we have subtracted 0.3 from the
normalized flux for display purposes), because that is the correct 
identification for its narrow P~Cygni component.}
\label{fig:hablue}
\end{figure}

The narrower H$\alpha$ profile of SN~2006tf is an obvious clue that
its blast-wave speed is significantly slower than that of SN~2006gy.
If we artificially stretch the velocity scale of the day 41 H$\alpha$
profile of SN~2006tf by a factor of 2, shown by the dotted line in
Figure~\ref{fig:ha06gy}, then the red wing of the line matches that of
SN~2006gy almost perfectly.  (A comparison on the blue side does not
work because of P~Cygni absorption in SN~2006gy.)  This provides a
strong case that the blast-wave speed in SN~2006tf is about half that
of SN~2006gy.  Therefore, we adopt 2,000 km s$^{-1}$ as the
characteristic blast-wave speed for SN~2006tf, defining the speed at
which the CSM is swept up by the forward shock. This also matches the
full width at half-maximum intensity (FWHM) of the intermediate
component of the H$\alpha$ line in SN~2006tf.  Judging from the
constant red wing (Fig.~\ref{fig:hanorm}), this speed remains
unchanged for at least the first 100~d, during a time when the SN is
radiating away $\sim$10$^{51}$ ergs in visual light.

\subsection{Underlying Broad Component}

As noted earlier, comparing the H$\alpha$ profiles at all four epochs
(Fig.~\ref{fig:hanorm}) shows that the red side of the line changes
very little, but there is extra emission that develops with increasing
strength on the broad blueshifted wing.  One would normally attribute
the apparent flux deficit in the red wing to high optical depth that
blocks the far side of the SN, but this hypothesis predicts a trend
opposite to the one seen; i.e., if the asymmetry were due to high
optical depths alone, then one would normally expect the broad
emission to become {\it more} symmetric with time.

Figure~\ref{fig:hablue} investigates the blueshifted wings of
broad-line profiles in more detail.  The profile of this ``extra''
emission on days 66 and 95 is illustrated in Figure~\ref{fig:hablue}
by subtracting the symmetric day 41 profile from the latter two
epochs.  The residual emission shows a broad component to H$\alpha$ at
blueshifted velocities, with its fastest speeds reaching about $-$7500
km s$^{-1}$, and little or no corresponding emission on the red side
of the line.

Two possible explanations for this fast blueshifted emission are that
it is due to electron scattering or to rapidly moving material.  If it
were due primarily to electron scattering, though, we should see a
component of comparable strength on the red side of the line as well.

In fact, the explanation for this fast blueshifted H$\alpha$ emission
is not likely to be electron scattering alone, because we also see
evidence for fast-moving blueshifted material in {\it absorption} in
some other species.  The bottom panel of Figure~\ref{fig:hablue} shows
broad profiles of O~{\sc i} $\lambda$7774 and He~{\sc i}
$\lambda$5876,\footnote{We have labeled this feature as He~{\sc i}
$\lambda$5876 because that places its narrow P~Cygni feature at the
correct velocity, but the underlying broad feature could also
plausibly be identified as Na~{\sc i} $\lambda$5895.} both of which
indicate weak absorption features with ill-defined blue absorption
edges out to about $-$7000 or $-$8000 km s$^{-1}$, matching the speeds
at which we see enhanced blueshifted H$\alpha$ emission.  We also note
that the relative strength of the broad emission component of He~{\sc
i} $\lambda$5876 increases with time.

Since the speeds up to about 7500 km s$^{-1}$ are well in excess of
the blast-wave speed we derived in the previous section, the broad
emission probably arises in gas that has not yet reached the reverse
shock of the CSM interaction region.  This expanding material just
interior to the reverse shock may be ionized by the
backward-propagating radiation from the post-shock gas (Chevalier \&
Fransson 1994).  The severe asymmetry of the line means that we are
only seeing the near side of the ejecta because of dust formation,
high optical depths in the ejecta, or global asymmetry.

The fact that these broad features are also seen in blueshifted {\it
absorption} in He~{\sc i} $\lambda$5876 and O~{\sc i} $\lambda$7774
provides a critical clue to their origin and to the physical
processes at work in this SN.  To be moving this fast, the material
must have not yet reached the reverse shock, but to be seen in
absorption, it must also have a background light source.  Thus, even as
late as day 95, some central source (i.e., the inner SN ejecta?) still
contributes a non-negligible fraction of the continuum luminosity.

What constitutes a ``non-negligible'' fraction in this case?  The
background continuum source must contribute at least 10\% of the
luminosity at red wavelengths, because that is roughly the depth of
the absorption in O~{\sc i} $\lambda$7774.  This requires the source
in question to have a luminosity of roughly
$8 \times 10^8$~L$_{\odot}$ (or $M_R \approx -17.4$ mag) on day 95, which
is about 1 mag more luminous than a normal SN~II-P at that same time
and even somewhat more luminous than a ``hypernova'' like SN~1998bw at
that epoch (Galama et al.\ 1998).  Thus, we must speculate that either
the underlying SN ejecta of SN~2006tf were unusually luminous
independent of its strong CSM interaction, or that perhaps the strong
CSM interaction has somehow rejuvenated the underlying ejecta with
inward-propagating radiation.  The alternative is that we would need
the 7500 km s$^{-1}$ to get out ahead of the 2000 km s$^{-1}$ shell,
which seems unlikely without radical departures from spherical
symmetry.

In these broad features, we have another clue that SN~2006tf is a more
luminous analog of SN~1988Z, which also showed underlying broad
components of H$\alpha$ in the SN ejecta (Stathakis \& Sadler 1991;
Turatto et al.\ 1993).  In the case of SN~1988Z, the underlying broad
H$\alpha$ component was also asymmetric, although not as severely as
that of SN~2006tf, and it also increased in prominence with time (up
to about day 115) and then faded again.  In principle, though, the
underlying SN ejecta would be easier to see in SN~1988Z because the
total SN luminosity was lower than in SN~2006tf.

\begin{figure}[ht]
\epsscale{1.0}
\plotone{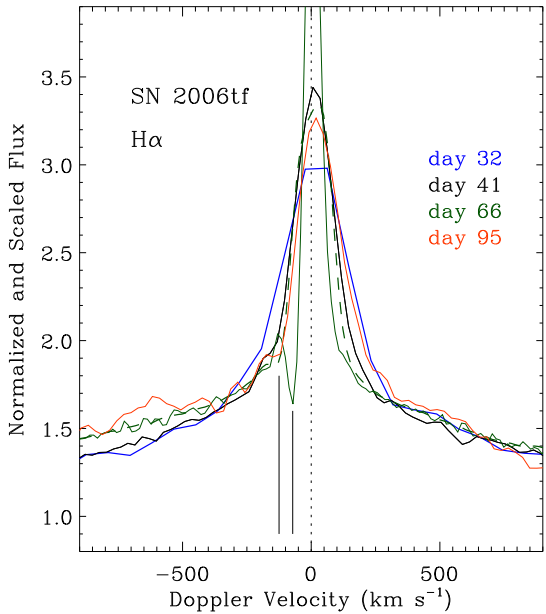}
\caption{Same as Figure~\ref{fig:hanorm}, but highlighting the narrow
H$\alpha$ component.  The scale has been adjusted so that the broad
wings of H$\alpha$ overlap.  The day 66 spectrum (green) was obtained
with DEIMOS using a higher resolution than the other spectra; it is
shown at the observed resolution (solid green), as well as smoothed
(dashed green) to match the resolution on days 41 (black) and 95
(orange).  At degraded resolution, the H$\alpha$ profile on day 66 is
nearly identical to that on days 41 and 95, suggesting little or no change 
in the strength of the narrow component or its narrow P Cygni absorption.
The day 32 spectrum (blue) was obtained with even lower resolution.
The two vertical dashes mark the velocity of the P Cygni trough at
$-$72 km s$^{-1}$ and its blue edge at $-$125 km s$^{-1}$.}
\label{fig:hanarrow}
\end{figure}

\subsection{Narrow H$\alpha$ and H$\beta$ Emission Profiles}

Our spectra at the four epochs are obtained with different spectral
resolution (see Table 2), and this may be the primary reason for
apparent differences in the narrow H$\alpha$ profile
(Fig.~\ref{fig:hanarrow}).\footnote{However, we do not believe that
the lower spectral resolution on day 64 fully accounts for the weaker
narrow absorption components of Fe~{\sc ii} and other species, as
compared to SNe~1994W and 2006gy in Figure~\ref{fig:allspec}, because
the spectrum with much higher resolution obtained 2 days later does not
show strong narrow absorptin features either.}  Only one epoch (day
66) had sufficiently high dispersion to fully resolve the line
profile.  It shows a strong narrow emission component with a width of
$\sim$150 km s$^{-1}$ in both H$\alpha$ and H$\beta$ at the point
where it meets the underlying broader base.  This is an underestimate
of the emission line's intrinsic full width near zero intensity (FWZI)
because a narrow P Cygni absorption feature is also seen in both
lines, which partially absorbs the blue side of the line.  Blueshifted
absorption is stronger in H$\beta$ than in H$\alpha$, which was true
for the broad components as well (Fig.~\ref{fig:hahb}).

\begin{figure}
\epsscale{1.0}
\plotone{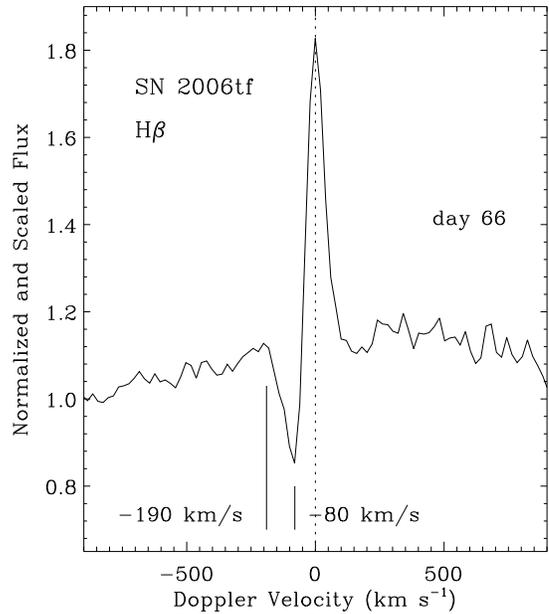}
\caption{Narrow component of the H$\beta$ profile on day 66 from the
high-resolution DEIMOS spectrum.  This gives the best estimate of the
progenitor's pre-shock wind speed of 190 km s$^{-1}$ from the blue
edge of the P Cygni absorption component. The minimum of the
blueshifted absorption is at $-$80 km s$^{-1}$.}\label{fig:hbnarrow}
\end{figure}

The pre-shock CSM speed is best seen in the blueshifted P~Cygni
absorption feature of H$\beta$ on day 66 (Fig.~\ref{fig:hbnarrow}),
which has a blue edge indicating a progenitor wind speed of 190
km~s$^{-1}$.  This is, again, similar to the case of SN~2006gy, where
the blue edge of the narrow P~Cygni absorption indicated a wind speed
of roughly 230 km s$^{-1}$ (Smith et al.\ 2007).  As noted in that
paper, such wind speeds are much too fast for red supergiants and much
too slow for Wolf-Rayet (WR) stars, but they are typical of blue
supergiants and especially LBVs.  As with SN~2006gy, this is likely to
be a critical clue to the nature of the star that exploded and its
immediate pre-SN mass loss that gave rise to its dense CSM.

Despite the differences in spectral resolution, we find no clear evidence
that the narrow H$\alpha$ profile is changing with time.  When we
degrade the spectral resolution of the day 66 spectrum (the {\it
dashed} green profile in Fig.~\ref{fig:hanarrow}) to match the
resolution on days 41 and 95, the narrow H$\alpha$ emission component
has the same profile shape and the P~Cygni absorption feature is no
longer seen.  This suggests that the same narrow P~Cygni absorption
may actually be present on days 41 and 95 as well, but is not resolved
in those spectra.  The first epoch on day 32 has even lower spectral
resolution, so a similar comparison is less clear.  There is some
suggestion of persistent narrow P~Cygni absorption in H$\beta$ at all
epochs (Fig.~\ref{fig:hahb}).

\begin{figure*}
\epsscale{0.9}
\plotone{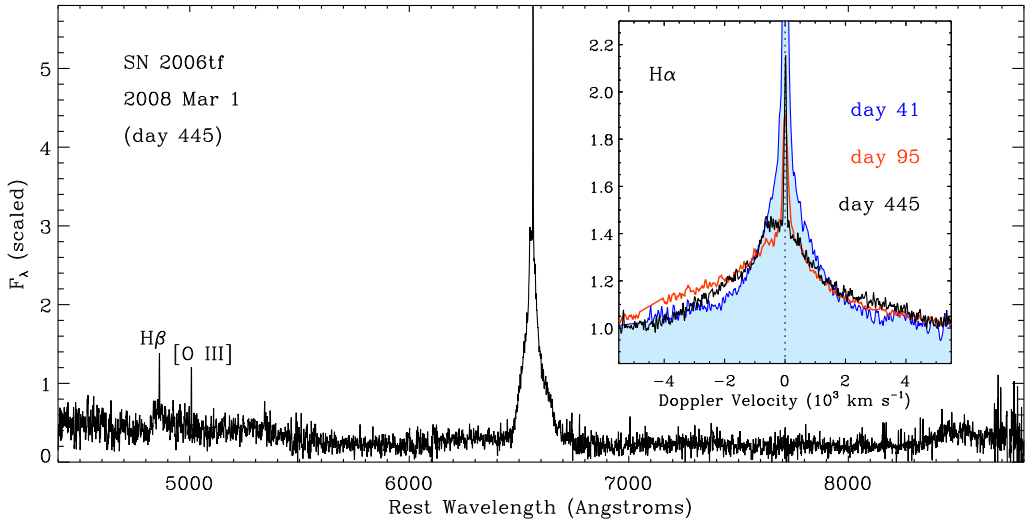}
\caption{The late-time spectrum of SN~2006tf, obtained on 2008 Mar.~1,
day 445 after discovery.  The inset shows the H$\alpha$ profile on day
445, scaled arbitrarily and compared to the day 41 (shaded blue) and
day 95 (orange) spectra from Figure~\ref{fig:hanorm}.  Narrow H$\beta$
and [O~{\sc iii}] $\lambda$5007 are real and may signify an underlying
H~{\sc ii} region, but there are many noise spikes redward of the
broad near-infrared Ca~{\sc ii} emission.}
\label{fig:mar08}
\end{figure*}

The strength of the narrow H$\alpha$ emission component relative to
the underlying broad component changes slowly but steadily.  If we
measure the strength of the narrow emission compared to the total line
flux including the broader emission (listed as ``N/T'' in Table 2), we
find that the narrow emission contributes roughly 17\%, 11\%, 8\%, and
then 3\% of the total H$\alpha$ emission on days 32, 41, 66, and 95,
respectively (Table 2).  As late as day 445, this fraction remains at
about 3--4\% (see below).  Recognition that the narrow component makes
a relatively minor contribution to the total line luminosity will be
important in \S 5.5, where we discuss the behavior of the integrated
line luminosities and equivalent widths.

\subsection{Late-Time H$\alpha$ Emission}

Our late-time spectrum taken on day 445 is shown in
Figure~\ref{fig:mar08}.  This is the first epoch at which we see
evidence that the overall character of the spectrum has changed
significantly.  The continuum emission is very weak, which is the
reason for the very large equivalent width in Table 2.  Besides
H$\alpha$, faint emission from other Balmer lines and the Ca~{\sc ii}
near-infrared triplet at 8500 \AA\ can also be seen.  The H$\alpha$/H$\beta$
flux ratio has now climbed to $\ga$11.6, indicating that it is
dominated by collisional excitation rather than recombination (Raymond
et al.\ 2007).

The broad components out to 7500 km s$^{-1}$ (Fig.~\ref{fig:hablue})
are no longer seen.  This suggests that the material responsible for
these features had reached the reverse shock by day 445.  The absence
of absorption, in particular, indicates that the power source of their
underlying continuum had faded by this time as well, as expected if
that continuum source was the underlying SN ejecta.

The spectrum is dominated by a strong H$\alpha$ emission line with a similar
intermediate width as before, as well as a lingering unresolved narrow
component that makes up a comparable fraction (about 3--4\%) of the total
emission as in the day 95 spectrum.  The narrow component is about a
factor of 2 narrower than before, though, with a FWHM of $\sim$80 km
s$^{-1}$, and reaching $\pm$80 km s$^{-1}$ at its base.  The narrow
P~Cygni absorption component is no longer present, even though the
late-time spectrum has higher resolution than our day 66 spectrum. The
luminosity of the narrow component is a factor of $\sim$6 fainter than on
day 95, with a luminosity of only $2.5 \times 10^{39}$ ergs s$^{-1}$.
This very narrow H$\alpha$ emission is likely to be at least partially
contaminated by a background H~{\sc ii} region, since narrow components
of H$\beta$ and [O~{\sc iii}] $\lambda$5007 are also seen
(Fig.~\ref{fig:mar08}).

The inset of Figure~\ref{fig:mar08} shows a detail of the H$\alpha$
line, compared to the early-time profiles on days 41 and 95.  The
broader component of the line has a similar width as before, and shows
complex behavior.  At some velocities, the line wings match the
symmetric day 41 profile, while at other velocities the late-time line
wings are similar to those on day 95.  These are probably important 
clues to the optical depth and geometry as the SN
evolves.

One very interesting quality of the late-time H$\alpha$ profile is the
pronounced asymmetry at relatively low velocities of $\pm$1000 km
s$^{-1}$.  Within this range, the intermediate-width component is very
asymmetric and blueshifted as compared to the symmetric day 41
profile, with a sharp edge on the blue side where it meets the line
wing.

An intriguing possibility is that the net blueshift of the line may be
due to dust formation in the dense post-shock cooling shell, analogous
to the well-established case of the peculiar SN~Ib 2006jc (Smith et
al.\ 2008a), and possibly also the SNe~IIn 2005ip (R.\ Chevalier 2008;
priv.\ comm.) and 1998S (Pozzo et al.\ 2004).  The optical depth would
be highest at the limbs of that shell, favoring the extinction of
material at low redshifted velocities and speeds near zero.  If dust
has formed in the post-shock region of SN~2006tf, we would predict
excess emission in near-infrared bands, analogous to the case of
SN~2006jc, and possibly also signs of increased extinction.  In that
case, the fact that dust formed much later in SN~2006tf (days 200--400
instead of by day 50 in SN~2006jc) can be understood as a consequence
of the fact that SN~2006jc faded much more rapidly than SN~2006tf, and
was much less luminous overall.  Dust formation could also mean that
the late-time continuum we measure is an underestimate.

\begin{figure}[!ht]
\epsscale{1.1}
\plotone{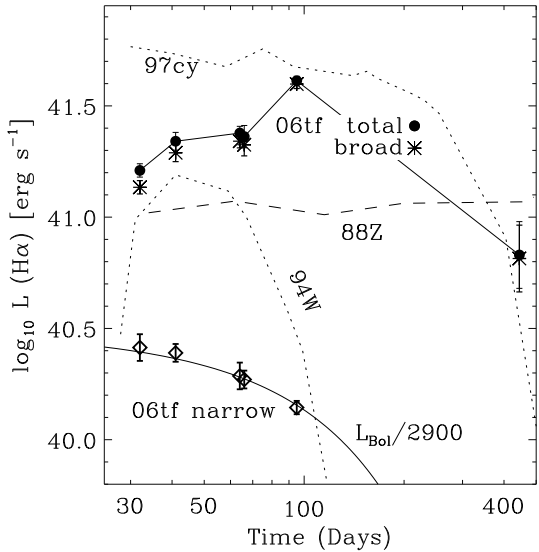}
\caption{The intrinsic luminosity of the H$\alpha$ emission line in
SN~2006tf, showing the total flux (solid dots), the
broad/intermediate-width component from the post-shock gas
(asterisks), and the narrow component from the unshocked CSM
(diamonds), from Table 2.  The decline rate of the narrow component of
H$\alpha$ in SN~2006tf follows the decline rate of the continuum
luminosity (the model diffusion curve from Fig.~\ref{fig:diffusion}
divided by 2900; solid curve).  For comparison, we also show the total
H$\alpha$ luminosities of SN~1994W (dotted line) from Chugai et al.\
(2004), and SN~1988Z (dashed line) from Stathakis \& Sadler (1991) and
Turatto et al.\ (1993), as well as SN~1997cy. [These SN~1997cy
measurements were taken from Fig.~11 of Pastorello et al.\ (2002).
A.\ Pastorello (2008, private comm.) informed us that they were
made from spectra presented by Turatto et al.\ (2000).]}
\label{fig:lum}
\end{figure}

\subsection{Line Luminosities and Equivalent Widths}

\begin{figure}[!ht]
\epsscale{1.1}
\plotone{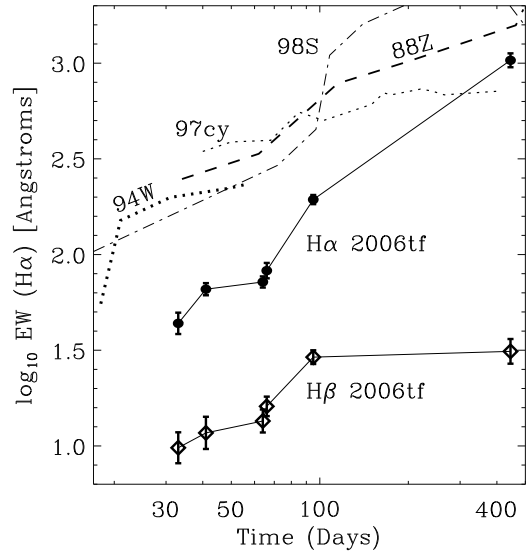}
\caption{The equivalent widths of H$\alpha$ and H$\beta$ emission with time,
measured in our spectra of SN~2006tf (emission-line equivalent widths
are positive and include narrow, broad, and intermediate-width
components).  Evolution of the H$\alpha$ equivalent width is shown for
several other representative SNe~IIn discussed in the text.  Except
for SN~1998S (Leonard et al.\ 2000), most objects do not have
published H$\alpha$ equivalent widths, so we derived them from $R$-band
photometry and published H$\alpha$ line fluxes from several sources
cited in the text.}\label{fig:ew}
\end{figure}

Figures~\ref{fig:lum} and \ref{fig:ew} show how the strength of Balmer
emission changes with time.  We measured the equivalent widths of the
total H$\alpha$ and H$\beta$ emission at the epochs for which we have
spectra, and used the red continuum flux inferred from the light curve
to derive the total H$\alpha$ line flux.\footnote{However, the
H$\alpha$ luminosity on day 445 was measured directly from the
flux-calibrated spectrum, although that flux was uncertain by roughly
50\% because of possible light cirrus clouds.} We also measured the
H$\alpha$/H$\beta$ flux ratio and the fractional contribution of the
narrow CSM emission components to the total H$\alpha$ flux in our
spectra. These are listed in Table 2.  Uncertainties in the H$\alpha$
flux are typically 5--8\%, depending on the noise in the continuum.

The luminosity of the narrow component declines slowly, roughly
following the same decline rate as the continuum luminosity inferred
from the light curve (Fig.~\ref{fig:lum} shows the same diffusion
model curve from Fig.~\ref{fig:diffusion} divided by a factor of 2900
for comparison).  Thus, the narrow H$\alpha$ emission represents a
constant 0.034\% of the bolometric luminosity, and its effective
equivalent width would be constant if it were measured relative to the
underlying continuum.  This implies that the narrow emission from the
pre-shock CSM is radiatively excited by the same continuum source that
powers the bolometric luminosity, but {\it not} the same source that
powers the broad post-shock H$\alpha$ emission.  If this is Case B
recombination emission, the mass of H gas needed to create the
H$\alpha$ line can be expressed as $M_{{\rm H}\alpha} = 11.4$
M$_{\odot}$ ($L_{{\rm H}\alpha}$/$n_e$), where $L_{{\rm H}\alpha}$ is
expressed in L$_{\odot}$ and $n_e$ is the characteristic electron
density in the CSM in cm$^{-3}$.  For the narrow H$\alpha$ component
in SN~2006tf, the value of $L_{{\rm H}\alpha} \approx 5 \times 10^6$
L$_{\odot}$ at early times implies a very large mass of pre-shock CSM
--- even for very high CSM densities, the total mass required is
$\sim$6~M$_{\odot}$ ($n_e$/10$^7$ cm$^{-3}$)$^{-1}$.  If this emitting
material resides within a radius of $\sim$10$^{16}$ cm, then the
average density is $\sim$10$^8$ cm$^{-3}$, implying that the
progenitor's wind had a mass-loss rate of $\sim$0.1 M$_{\odot}$
yr$^{-1}$ for many decades before the SN, and prior to the even more
extreme mass loss in the few years before explosion.

\begin{figure*}[ht]
\epsscale{0.75}
\plotone{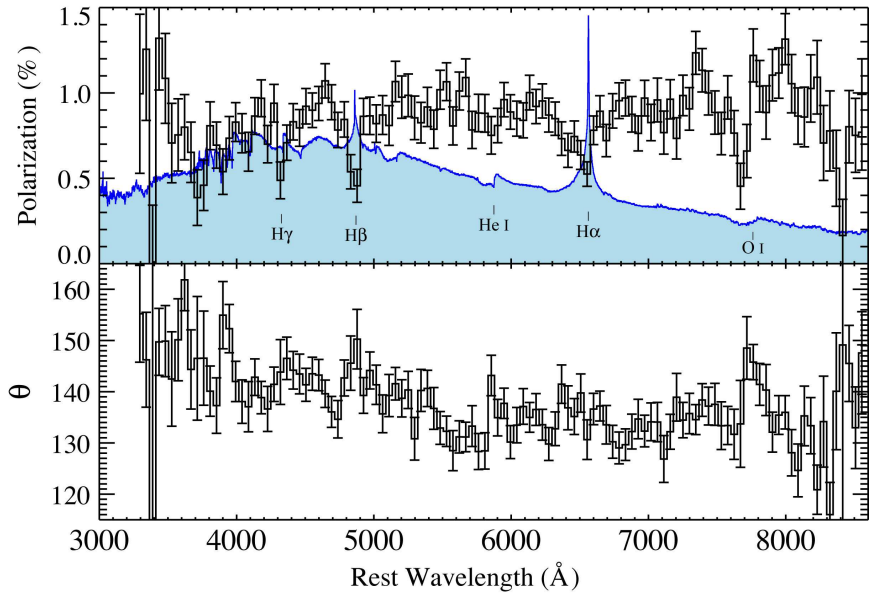}
\caption{Spectropolarimetry of SN~2006tf on day 64.  The top panel
shows the observed polarization (estimated using a rotated Stokes
parameter) in black with error bars, binned to 50 observed \AA\
pixel$^{-1}$ for clarity.  Overplotted in blue (shaded gray in the
printed edition) is the total-flux spectrum from that date for
reference.  The continuum polarization between the lines is about
0.9\%, with decreases in polarization visible at the wavelengths of
the Balmer emission lines and especially O~{\sc i} $\lambda$7774.  The
bottom panel shows the observed position angle of the polarization.}
\label{fig:specpol}
\end{figure*}

The broad/intermediate-width component (and hence, the total H$\alpha$
line luminosity) exhibits very different behavior from that seen in
the narrow CSM lines (Fig.~\ref{fig:lum}).  Instead of fading with the
continuum luminosity, the H$\alpha$ luminosity of the post-shock gas
rises dramatically as the SN fades.  An important clue to this
behavior is that the nature of the intermediate-width Balmer emission
changes dramatically during this time (see Table 2): at early times,
the H$\alpha$/H$\beta$ ratio matches the value one expects for
recombination emission, suggesting that the heating of the post-shock
gas emitting H$\alpha$ is actually dominated by photoionization
heating.  This changes with time, however, as the ratio climbs
to H$\alpha$/H$\beta$ $>$10, indicative of pure collisional
excitation.  It can also be seen in the equivalent-width behavior
(Fig.~\ref{fig:ew}), where H$\alpha$ continues to rise, while H$\beta$
levels off.

This suggests that as time proceeds, SN~2006tf transitions from an
optically thick regime where radiative heating dominates the emission,
to a regime where collisional heating by ongoing CSM interaction
dominates.  In some sense, this will be true for any strongly
interacting SN~IIn, because the highest optical depths are usually
encountered at early times.  One clear manifestation of such behavior
is an increasing line-to-continuum ratio, measured as the equivalent
width.  Figure~\ref{fig:ew} compares the H$\alpha$ equivalent width of
SN~2006tf to that of several other SNe.  All SNe~IIn shown here have
H$\alpha$ equivalent widths that increase with time as the densities
drop and the underlying continuum fades.  However, it is clear that
even though the H$\alpha$ luminosity of SN~2006tf is not too different
from some of these other SNe (Fig.~\ref{fig:lum}), its H$\alpha$
equivalent width is systematically less than all others during the
main peak of its light curve, and it does not catch up until late
times (Fig.~\ref{fig:ew}).  This discrepancy would be even more stark
if we accounted for the likely fact that the explosion date should be
at least 30~d before the discovery date, shifting the SN~2006tf data
to the right in Figure~\ref{fig:ew}.  This is a sign that the
continuum luminosity contributed by photon diffusion from optically
thick material is much stronger in SN~2006tf than in the fainter
SNe~IIn, and that this component persists much longer.  At early
times, the fraction of the bolometric luminosity contributed by the
H$\alpha$ line, $L_{{\rm H}\alpha}/L_{\rm Bol}$, is about 0.4\%, rising to
about 2.6\% at late times.  Corresponding numbers for the other
SNe~IIn in Figure~\ref{fig:ew} are $L_{{\rm H}\alpha}/L_c \approx 2$\%
at early times and about the same as the of SN 2006tf at late times.

The different behavior with time in SN~2006tf suggests that it would
be difficult to fully account for its high luminosity with a model
like the one applied to most other SNe~IIn, where H$\alpha$ and
continuum emission cool the post-shock gas in quasi-steady-state.  The
difference in the case of SN~2006tf is that the extra continuum
luminosity may be due to delayed photon diffusion from an opaque
shocked shell, as discussed earlier.  Without a two-component model
like this, it is difficult to see why the continuum luminosity of
SN~2006tf would be a factor of 4--10 higher than that of SNe~1994W and 1988Z
even though its H$\alpha$ luminosity is only 1.5--2 times larger.

Eventually, as the radius increases and the density drops, SN~2006tf
becomes dominated by direct radiation from ongoing optically thin CSM
interaction, very much like the late-time behavior of SN~1988Z and
other SNe~IIn. However, the drop in the total H$\alpha$ luminosity at
late times suggests that the progenitor mass-loss rate of SN~2006tf
was not constant, but began to rise as the time of core collapse
approached (see \S 7).

\section{Spectropolarimetry}

The spectropolarimetry of SN 2006tf on day 64 is plotted in
Figure~\ref{fig:specpol}.  The continuum is polarized at about the
0.9\% level, with several depressions in the polarization at the
locations of the strong Balmer emission lines and O~{\sc i}
$\lambda$7774.  For H$\alpha$ and O~{\sc i} especially, the
depolarization seems to be primarily associated with the broad
blueshifted components discussed in \S 5.2 and
Figure~\ref{fig:hablue}.  The differing polarization of the lines and
continuum indicates that at least some of the observed polarization
must be intrinsic to the supernova and not simply due to ISP.  We
integrated the observed polarization over the spectral range
5050--5950~\AA \ in the rest frame of the supernova to simulate a
rest-frame $V$-band observation and obtained a value of $P_V = 0.91
\pm 0.03$\% for the continuum polarization at a position angle of
$\theta = 135\fdg4 \pm 0\fdg8$.

The few SNe~IIn studied polarimetrically in the past have hinted that
the objects as a class exhibit high polarizations and hence large
asymmetries.  The first evidence that this might be true came from
broad-band imaging polarimetric observations of SN~1994Y about 245~d
after discovery (Wang et al.\ 1996).  The $R$-band polarization for
that object (dominated by H$\alpha$) differed by more than 1.5\%
(defined as the difference in the Stokes parameters $q$ and $u$ added
in quadrature) from that measured in the $B$ and $V$ bands, which were
dominated by continuum emission.  Wang et al.\ (1996) concluded that
SN~1994Y had significant intrinsic polarization, but without
spectropolarimetry the interpretation was unclear.

\begin{deluxetable*}{lcccccccc}[!ht]\tabletypesize{\scriptsize}
\tablecaption{Some Basic Physical Properties of
SN~2006\lowercase{tf}\tablenotemark{a} in a CSM Interaction Scenario}
\tablewidth{0pt}
\tablehead{
  \colhead{Day} &\colhead{L$_{\rm Bol}$} &\colhead{T$_{BB}$} &\colhead{$R_{BB}$} 
  &\colhead{$R_{\rm shell}$}  &\colhead{$\zeta$}  &\colhead{$w$}
  &\colhead{$\dot{M}_{\rm CSM}$} &\colhead{$t_{\dot{M}}$} \\
  \colhead{} &\colhead{(10$^9$ L$_{\odot}$)} &\colhead{(K)} &\colhead{(10$^{15}$ cm)} 
  &\colhead{(10$^{15}$ cm)} &\colhead{} &\colhead{(10$^{18}$ g cm$^{-1}$)} 
  &\colhead{(M$_{\odot}$ yr$^{-1}$)} &\colhead{(yr)} 
}
\startdata
32   &13.6  &7800   &4.52   &4.56  &0.98    &13.6  &4.1  &$-$7.6  \\
41   &12.7  &7500   &4.72   &4.72  &1.00    &12.7  &3.8  &$-$7.9  \\
64   &10.5  &6800   &5.22   &5.12  &1.04    &10.5  &3.1  &$-$8.6  \\
66   &10.3  &6800   &5.18   &5.16  &1.02    &10.3  &3.1  &$-$8.6  \\
95   &7.9   &6300   &5.26   &5.66  &0.86    &7.9   &2.3  &$-$9.5  \\
445  &0.64  &[6300] &[1.5]  &11.7  &[0.016] &0.64  &0.2  &$-$20   \\
\enddata
\tablenotetext{a}{$R_{\rm shell}$ and properties in columns to its right
are derived assuming a constant shell expansion speed of 2000 km
s$^{-1}$.  Highly uncertain values are in square brackets}
\end{deluxetable*}

Subsequently, spectropolarimetric sequences of two SNe~IIn have
appeared in the literature.  SN~1998S was observed within a week of
explosion (Leonard et al.\ 2000) and twice at later epochs (Wang et
al.\ 2001).  Hoffman et al.\ (2008) recently presented three epochs of
spectropolarimetry of SN~1997eg ranging from 16 to 93~d after
discovery.  The initial continuum polarization was observed to be
large in both objects ($P \approx 2.0$\% in SN~1998S and $P \approx
2.3$\% in SN 1997eg), with modulations larger than 1.5\% across the
strongest emission lines.  Both objects also showed time-variable
polarization in the continuum and lines, with the polarization
changing by more than a percent in the $q-u$ plane.

Despite the high data quality for these two SNe, interpretation has
been complicated.  Observed continuum polarization can be attributed
to both ISP by dust along the line of sight and intrinsic continuum
polarization of the SN due to electron scattering.  Scattering by
lines is believed to be intrinsically depolarizing, but if the
line-scattering material is not distributed spherically, it may
produce net polarization when the spatially unresolved SN is observed
from afar.  Finally, as discussed above, emission-line profiles show
several components representing material with potentially different
spatial distributions.  The earliest epoch of data on SN~1998S
(Leonard et al.\ 2000) showed separate polarization modulations in
both narrow and broad lines, prohibiting any interpretation that
allowed both components to be completely depolarizing.  Hoffman et
al.\ (2008) found that the polarization of strong lines (H$\alpha$,
H$\beta$, and He~{\sc i} $\lambda$5876) in SN~1997eg showed ``loops''
when plotted in the Stokes $q-u$ plane.  They attributed this to the
differing geometries of narrow- and broad-line material, and thus to a
mismatch in symmetry axes of SN ejecta and the CSM.

Our observations of SN~2006tf do not have sufficient signal-to-noise
ratio to disentangle all these effects, especially from a single epoch
of data, but we can make useful comparisons to SNe~1998S and 1997eg.
The first point to note is that the observed continuum polarization of
SN~2006tf is lower than that of the other two objects, both of which
showed polarizations of at least 2\% at early times.  However, both
objects showed significant ISP, making any estimate of the intrinsic
continuum polarization difficult and model dependent.  Also, the
continuum polarization changed over time, with that of SN~1997eg
declining to $\sim$1.5\% by day 93.  Our SN~2006tf observations were
taken at least 64~d after explosion, so one possibility is that the
polarization was higher at earlier times, but the optical depth to
electron scattering declined with time, perhaps as a result of
recombination or decreasing densities in the ejecta.

Potential evidence that the intrinsic continuum polarization is
actually lower in SN~2006tf than in the other two objects is that the
depolarization of line features is significantly smaller.  As
mentioned above, the line features in the other objects were stronger
than 1.5\%.  In SN~2006tf, the modulation at H$\alpha$ is only about
0.4\%.  Quantitatively, the Stokes parameters ($q$,$u$) integrated
over 6555$-$6575~\AA\ (the core of the narrow H$\alpha$ line) are
($-$0.27, $-$0.60) with an uncertainty of $\pm$0.09\% for each.  This
value represents a difference of 0.41\% in polarization (Stokes
parameters added in quadrature) from the $V$-band value quoted above,
corresponding to $\sim$1/4 of the polarization in the other objects.
Overall, then, our general conclusion is that SN~2006tf is only
moderately polarized at late times, showing no clear sign for extreme
large-scale asymmetry in the CSM interaction region.

\begin{figure}
\epsscale{1.0}
\plotone{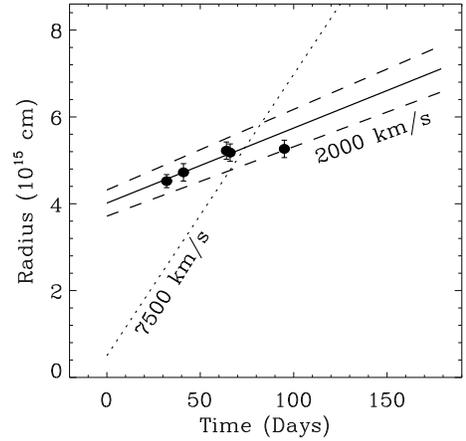}
\caption{Filled dots show the derived black-body radius from Table 3,
  appropriate for the luminosity and temperature derived from the
  spectra in Figure~\ref{fig:allspec}.  The solid line shows the
  radius for the observed nominal shock speed of 2000 km s$^{-1}$, and
  the two dashed lines show representative inner and outer radii where
  the thickness of the post-shock shell is $\sim$15\% of the radius.
  The dotted line shows radii for an expansion speed of 7500 km
  s$^{-1}$, where SN ejecta can almost reach the reverse shock by day
  64 when the fast blueshifted emission is seen at those speeds.}
\label{fig:radius}
\end{figure}

\section{Basic Physical Properties of SN~2006\lowercase{tf}}

In the preceding sections, we have described the basic energy demands,
overall spectral properties, kinematics from line profiles, and line
luminosities that are important clues to the nature and evolution of
SN~2006tf.  With these in hand, we can now quantify some of the
fundamental physical properties of the SN, listed in Table 3 for each
day since discovery when we have spectra.  Column 2 lists the
bolometric luminosity (Fig.~\ref{fig:diffusion}) derived from $R$-band
photometry, and column 3 lists the characterstic black-body
temperatures for the continuum shapes in Figure~\ref{fig:allspec}.
These are used to calculate the corresponding radius that a black body
would have, listed as $R_{BB}$ in column 4. In column 5, $R_{\rm
shell}$ is the radius of the shell expanding at a constant speed of
2000 km s$^{-1}$, starting from a large initial radius suitable for
$R_{BB}$ at the early epochs.  These two quantities are also plotted
in Figure~\ref{fig:radius}.  Column 6 lists the corresponding dilution
factor, $\zeta = R_{\rm BB}^2/R_{\rm shell}^2$, which provides an
estimate of the effective optical depth or geometric covering factor
of the emitting regions of the shell.  This dilution factor is a key
component in the transition from optically thick to optically thin, as
we will discuss later.  The next two columns give the wind-density
parameter $w = \dot{M}_{\rm CSM}/V_{\rm CSM}$ and the progenitor
mass-loss rate $\dot{M}_{\rm CSM}$.  These are the values needed to
account for the observed bolometric luminosity of SN~2006tf in a
simple CSM-interaction model with 100\% efficiency, where the observed
luminosity is emitted instantaneously by the post-shock gas,
calculated from the expression

\begin{equation}
\dot{M}_{\rm CSM} = 2 \ L \ \frac{V_{\rm CSM}}{(V_{\rm shell})^3},
\end{equation}

\noindent 
where $V_{\rm CSM} = 190$ km s$^{-1}$ is the observed pre-shock CSM
wind speed, and $V_{\rm shell} = 2000$ km s$^{-1}$ is the observed
constant speed of the post-shock shell.  Again, these provide an
estimate of the mass-loss rate that created the CSM being swept up to
produce the observed luminosity.  Since this assumes 100\% efficiency,
it may be an underestimate; any clumping or global asphericity will
raise the required values of $w$ and $\dot{M}_{\rm CSM}$.  Accounting
for possible broadening of the line wings by electron scattering would
raise $\dot{M}_{\rm CSM}$ even further because it would imply a
smaller value of $V_{\rm shell}$.  The last column lists the CSM flow
timescale $t_{\dot{M}} = R_{\rm shell}/V_{\rm CSM}$, which is the
number of years prior to core collapse when the progenitor had the
mass-loss rate in the previous column.

The quantities in the last two columns are of particular interest,
because they provide a roadmap for what the progenitor star was doing
in the decades before it died.  This indicates that the progenitor
underwent a sudden boost in mass loss in the decade just before
exploding.  A caveat is that the instantaneous mass-loss rate may not
have been quite as high as the peak values around 4 M$_{\odot}$,
because diffusion of radiation from an optically thick shocked shell
has the effect that the luminosity from earlier CSM interaction, when
the material was highly opaque, will effectively ``pile up'' and mimic
a larger instantaneous mass-loss rate.  Even allowing for this effect,
though, average mass-loss rates of order 2 M$_{\odot}$ yr$^{-1}$ are
needed in the decade before explosion.  Before that time (at $t$ minus
20 yr), the mass-loss rate was apparently 10 times less.

\begin{figure*}[t]
\epsscale{1.0}
\plotone{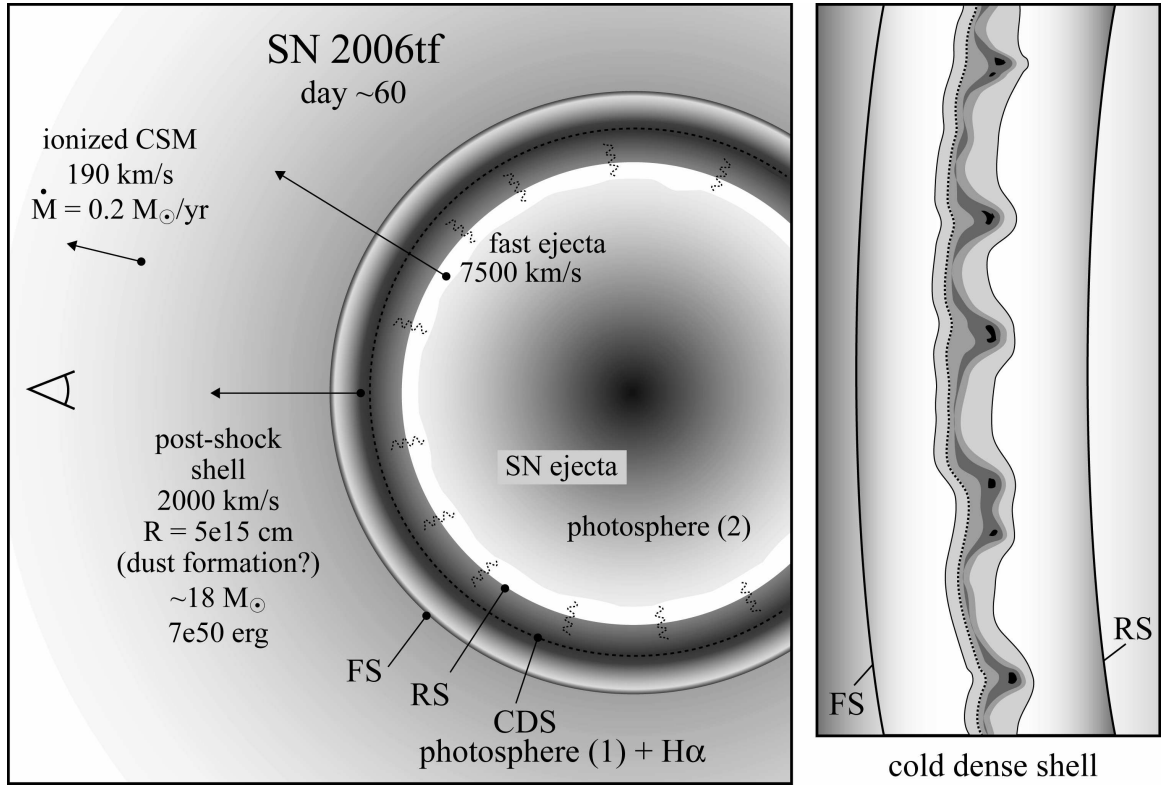}
\caption{Cartoon illustration of the components of SN~2006tf at about
   60~d after discovery, during the decline from the main light-curve
   peak.  The primary feature is the massive post-shock shell of gas,
   composed of the swept-up opaque pre-SN envelope around the star
   ejected in the decade before core collapse.  Most of the mass is in
   the cold dense shell (CDS), bounded by the forward shock (FS) and
   the reverse shock (RS).  Diffusion of radiation from this shocked
   shell produces the main continuum photosphere (1) and the
   intermediate-width component of H$\alpha$. This shell expands at
   constant speed into the pre-shock CSM (dense wind of the
   progenitor).  The interior of the shell is filled by freely
   expanding SN ejecta, the outermost parts of which are ionized by
   radiation (wavy lines) propagating inward from the reverse
   shock, exciting the broad He~{\sc i} and O~{\sc i} features seen in
   the spectrum.  There is also a second photosphere (2) in the SN
   ejecta, which is fainter than the main photosphere and can only be
   seen if the main shell thins or develops clumps as time proceeds.
   The {\it right panel} shows a more detailed depiction of the
   post-shock gas, including the clumpy structure that forms due to
   instabilities in the cold dense shell layer.  The dashed line here
   represents the photosphere at some arbitrary early time, working
   its way from left to right through the clumpy CDS as the SN
   expands.  When it reaches a dense clump, the recombination
   photosphere will proceed through that clump, but for the regions
   between clumps it will eventually break through, allowing an
   observer to see the underlying SN ejecta.}
\label{fig:sketch}
\end{figure*}

As we have noted elsewhere for SN~2006gy (Smith et al.\ 2007, 2008b;
Smith \& McCray 2007), the only stars known to be capable of producing
such extreme mass-loss rates of this order are LBVs during their giant
eruptions (Smith \& Owocki 2006), like the 19th-century eruption of
$\eta$ Carinae when the star ejected 12.5 M$_{\odot}$ in about a
decade (Smith et al.\ 2003).  Even the factor of 10 lower mass-loss
rate of 0.1--0.2 M$_{\odot}$~yr$^{-1}$ needed in the decades before
that for SN~2006tf is still 1000 times stronger than the maximum rate
that can be supplied by a line-driven wind of a massive star (Smith \&
Owocki 2006).  However, 0.1--0.2 M$_{\odot}$~yr$^{-1}$ is comparable
to the less extreme LBV eruption of P Cygni in 1600 AD (Smith \&
Hartigan 2006).  Gal-Yam et al.\ (2007) have also discussed LBVs as
potential SN~IIn progenitors for independent reasons.  If the stars
that exploded as SNe~2006tf and 2006gy were {\it not} actually LBVs,
then they have done a remarkably good job of impersonating the H-rich
composition, amount of mass ejected, and ejection speeds of giant LBV
eruptions.

\section{SUMMARY AND DISCUSSION}

\subsection{Summary of Observational Clues}

The dataset we have presented above includes many different insights
to the nature of SN~2006tf that weave a complex picture of a very
luminous SN with remarkably strong CSM interaction.
Figure~\ref{fig:sketch} shows an illustration of the essential structural 
elements of SN~2006tf following the hypothesis that it is a case of
extremely strong CSM interaction, including diffusion from an opaque
shocked shell, as mentioned many times throughout this paper.  We now
summarize the key observational clues of SN~2006tf presented in our study,
along with the main implications of each and how they might be
understood in a simplified model like that portrayed in
Figure~\ref{fig:sketch}.


1. The total energy radiated at visual wavelengths is at least $7
  \times 10^{50}$ ergs.  It could be substantially more depending on
  how soon after explosion SN~2006tf was discovered.  If momentum is
  conserved, then at best, roughly half the initial kinetic energy can
  be converted into radiation in a CSM-interaction model.  This
  suggests that the expanding shell should retain roughly an equal
  amount of kinetic energy.  Indeed, our estimated shell mass of
  18~M$_{\odot}$ expanding at 2000 km s$^{-1}$ does yield $7 \times
  10^{50}$ ergs. This requires a total initial explosion energy of at
  least $1.4 \times 10^{51}$ ergs (not including the kinetic energy of
  fast SN ejecta that have not yet reached the reverse shock).  In
  other words, the very large CSM mass we estimate is comparable to
  that which is needed to absorb momentum and decelerate the blast
  wave to only 2000 km s$^{-1}$.

2.  The decline rate of the optical light curve roughly matches that
    expected for the decay rate of $^{56}$Co, including the late-time
    photometry more than 1~yr after explosion
    (Fig.~\ref{fig:diffusion}).  The initial mass of $^{56}$Ni needed
    to provide such a high luminosity is about 4.5~M$_{\odot}$.  In
    that case, a pair-instability explosion would need to be invoked,
    as in the case of SN~2006gy if it was also powered by radioactive
    decay (Smith et al.\ 2007).  In our estimation, however, the
    strong signature of CSM interaction in the spectrum makes the
    radioactive-decay hypothesis less palatable for SN~2006tf.

3.  The decline rate of the optical light curve during its main peak
    is fit even better by a simple model of photon diffusion from an
    opaque shocked shell (Smith \& McCray 2007).  This model requires
    a shell mass of roughly 18~M$_{\odot}$ to account for the long
    diffusion time and high luminosity, which matches the large mass
    needed to decelerate the blast wave as noted in point (1) above.
    The initial radius of the opaque shocked shell is about 180 AU or
    $\sim 3 \times 10^{15}$ cm, close to the radius one might expect
    shortly before discovery (Fig.~\ref{fig:radius}).

4.  However, both radioactivity and this diffusion model fall short of
    the luminosity needed to power the late-time photometry of
    SN~2006tf (Fig.~\ref{fig:diffusion}).  The similar CSM-interaction
    model by Woosley et al.\ (2007), requiring roughly 25 M$_{\odot}$
    of CSM (but admittedly tailored for SN~2006gy instead of
    SN~2006tf), has the same difficulty (Fig.~\ref{fig:diffusion}).
    Thus, in any CSM-interaction model, we need to invoke additional,
    ongoing CSM interaction at large radii to explain the late-time
    behavior.  Indeed, such ongoing CSM interaction is evidenced by
    the strong H$\alpha$ emission at late times.  The progenitor
    mass-loss rate needed to account for the late-time luminosity
    through CSM interaction is about 0.2~M$_{\odot}$ yr$^{-1}$, about
    20 yr before explosion.  This is within a factor of 2 of the value
    required by the narrow CSM component (see next point).

5.  The narrow emission and P~Cygni absorption components of H$\alpha$
    and H$\beta$ reveal that the pre-shock CSM is expanding at roughly
    190 km s$^{-1}$.  This is an important clue to the nature of the
    progenitor star, because it makes a red supergiant or WR-star progenitor 
    seem unlikely, while the speed is comparable to those of LBVs and other
    blue supergiants.  The luminosity of the narrow H$\alpha$
    component implies a progenitor mass-loss rate at large radii
    outside the shock ($\sim$10$^{16}$ cm; ejected about 20 yr before
    core collapse) of at least 0.1~M$_{\odot}$ yr$^{-1}$.  This is
    close to the value needed to power the late-time luminosity (see
    previous point), and fully consistent within the uncertainty of the
    late-time luminosity estimate. This is also a factor of $\sim$10 lower
    than the necessary mass-loss rate in the decade just before core
    collapse, signifying a sharp boost in $\dot{M}$ immediately before
    the star's death.

6.  The intermediate-width component of the H$\alpha$ line arises
    mostly in a swept-up, dense, post-shock cooling shell expanding at
    a constant speed of $\sim$2000 km s$^{-1}$
    (Fig.~\ref{fig:sketch}).  This is the dominant speed of the
    forward shock plowing into the CSM.  This speed does not change
    perceptibly from day 32 onward.  Since the shell does not
    decelerate even though it is emitting almost 10$^{51}$ ergs, the
    shell must already be very massive by day 32, consistent with our
    estimates above.

7.  The nature of the Balmer emission changes with time.  At early
    times, the H$\alpha$/H$\beta$ flux ratio is consistent with
    recombination, whereas at late times, the H$\alpha$/H$\beta$ ratio
    rises to more than 10, suggesting that it becomes dominated by
    direct collisional excitation.

8.  Broad wings of H$\alpha$ may be due in part to electron
    scattering, but there also appears to be an underlying broad
    emission component, seen almost exclusively at blueshifted speeds
    up to about $-$7500 km s$^{-1}$ (Fig.~\ref{fig:hablue}).  This
    broad component appears sometime after day 41, is seen on days 64
    through 95, and disappears again at very late times.  We propose
    that this feature corresponds to the outermost parts of the SN
    ejecta that have almost reached the reverse shock (see
    Fig.~\ref{fig:sketch}).  Material traveling at this speed would,
    in fact, just about reach the radius of the reverse shock by this
    time after explosion (Fig.~\ref{fig:radius}).  The broad features
    are also seen in P Cygni absorption in He~{\sc i} $\lambda$5876
    and O~{\sc i} $\lambda$7774.  The absorption requires some
    additional background continuum light source, which is likely to
    be the diffusion of radiation from the inner SN ejecta deposited
    by shock energy or radioactive decay.  The luminosity required for
    the absorption strength implies that the underlying SN was
    overluminous as well, independent of CSM-interaction.

9.  A possible explanation for why the broad features are seen only
    from day 64 through 95 is that before that time, the shocked shell
    was highly opaque (the broad features reside interior to the
    reverse shock; Fig.~\ref{fig:sketch}).  Long after that time (by
    day 445), the SN ejecta luminosity has probably dropped far below
    that of the ongoing CSM-interaction region.

10. The luminosity of the intermediate-width component of H$\alpha$ is
    not correlated with the continuum luminosity of the SN
    (Fig.~\ref{fig:lum}).  It rises as the continuum luminosity
    fades. Compared to other SNe~IIn, the H$\alpha$ equivalent width
    is lower, but rises to similar values at late times more than 1 yr
    after explosion (Fig.~\ref{fig:ew}).  This is another clue that
    SN~2006tf has some additional source of continuum luminosity at
    early times, which is likely to be the slow diffusion of radiation
    from the massive swept-up opaque shell that mimics a normal
    H-recombination SN atmosphere, but at constant velocity.

11. The intermediate-width post-shock H$\alpha$ emission has
    pronounced asymmetry at late times, showing an asymmetric and
    blueshifted profile at velocities within roughly $\pm$1000 km
    s$^{-1}$.  This may hint that dust formation has occurred in the
    dense post-shock cooling shell, analogous to SN~2006jc (Smith et
    al.\ 2008a), blocking gas mainly at the limbs and partially at the
    back side of the shell (bottom and top of the shell in
    Fig.~\ref{fig:sketch}).

12. One epoch of spectropolarimetry of SN~2006tf reveals relatively
    mild continuum polarization of 0.9\% and also moderate
    depolarization of Balmer lines.  Both effects are weaker than in
    other SNe~IIn for which data are available, so we conclude that
    SN~2006tf shows no severe, global asymmetry.  Thus, the simplified
    spherical scenario we describe next is basically valid.

\subsection{Conclusion:  Optically Thick to Optically Thin}

Altogether, then, we conclude that available evidence requires a
two-component model for the power source of SN~2006tf.  These two
components correspond to different regimes of CSM interaction: (1)
shock interaction with extremely dense and opaque CSM with progenitor
mass-loss rates of order 1~M$_{\odot}$ yr$^{-1}$, where the material
has such high optical depth that the radiation generated by the shock
interaction must diffuse out slowly through the massive amount of
material after a delay corresponding to the diffusion timescale, and
(2) more conventional, optically thin CSM interaction, where the
luminosity generated by CSM interaction is able to escape without
significant delay.  As SN~2006tf expands and thins, it makes a
transition between these two regimes.

Exactly how and when does this transition occur?  In a normal SN~II-P,
the H-recombination photosphere recedes backward through the expanding
SN ejecta, maintaining a roughly constant $R_{\rm BB}$ during the
plateau phase as it passes through ejecta with a large gradient in
expansion speed.  The luminosity eventually drops sharply as the
photosphere recedes all the way back through the H layer and the SN
enters the nebular phase.  This cannot be the case in SN~2006tf
because the massive shell is expected to be geometrically thin and the
observed expansion velocity is constant, while we observe no sudden
transition akin to SNe~II-P.

We propose that in SN~2006tf, the transition from an optically thick
dense shell to the optically thin CSM interaction phase happened
gradually as a result of clumping in the cold dense shell.  The
contact discontinuity, where material piles up between the forward and
reverse shocks, will be severely Rayleigh-Taylor unstable and will not
be a homogeneous constant-density spherical shell
(Fig.~\ref{fig:sketch}, {\it right panel}). The size scale of density
fluctuations will be similar to the thickness of the dense shell,
which is geometrically very thin because it is relatively cold.  Thus,
the clumps that develop will have a typical size much smaller than the
radius of the shell.  The large number of small clumps across the
surface of the shell will remain optically thick as recombination
fronts move through them and their thermal energy is slowly radiated
away, but the less dense regions between clumps will become optically
thin more quickly, and the spacing between clumps will grow with time
until all the material in the shell eventually becomes optically thin.
Therefore, we would expect that the observed transition from
effectively optically thick to effectively optically thin can appear
to be a smooth transition, where the shrinking geometric covering
factor of the optically thick clumps will mimic a shrinking average
optical depth, while still allowing diffusion to be important for
energy leakage from individual thick clumps.  In effect, by looking
through the increasingly porous shell, an observer can see the underlying
SN ejecta even though the clumped material in the shell itself is
still optically thick.

This decrease in the effective covering factor of opaque clumps
(rather than an actual decrease in the radius) is the ``dilution
factor'' $\zeta$ in Table 3, needed to explain the gradual fading as
the shell expands.  Thus, diffusion of shock-deposited energy from an
opaque envelope dominates at early times (as in SN~2006gy; Smith \&
McCray 2007), but as time progresses, the subsequent ongoing CSM
interaction remains constant and eventually comes to dominate the
late-time luminosity and spectrum as the optically thick shell fades.
This is why the H$\alpha$/H$\beta$ flux ratio appears to rise steadily.

At late times, this opaque shocked shell will have radiated away the
bulk of its thermal energy and will cool significantly, heated only by
the ongoing shock energy.  The average densities in this
18~M$_{\odot}$ thin shell are extremely high --- roughly 10$^{11}$
cm$^{-3}$ if the shell thickness is $\sim$10\% of its radius.  With
the low luminosity and large radius at late times, it would not be
surprising if dust formation occurs in the opaque cool shell, for
which we see suggestive evidence in the late-time blueshifted
H$\alpha$ line profile.  We encourage near-infrared spectroscopy of
SN~2006tf in order to determine if it shows the expected
2.3~$\micron$ CO bandhead emission, as was seen in SN~1998S (Gerardy
et al.\ 2000; Fassia et al.\ 2001).

We might expect that this transition from thick to thin could, or
should, happen for any SN~IIn where the CSM close to the star is dense
enough to be opaque, and this is probably true for other luminous
SNe~IIn at early times (e.g., SN~1994W; Chugai et al.\ 2004).  The
remarkable thing about SN~2006tf is just how high that luminosity was,
how dense the CSM must have been, and how long the optically thick
phase persisted.  The wind density parameter, $w$, for SN~2006tf is
orders of magnitude higher than that of most other SNe~IIn.  This is
attributable to the very high mass-loss rate of the progenitor of
SN~2006tf, but also to the fact that its wind speed of 190 km s$^{-1}$
is relatively slow (SN~1994W, for example, apparently had a CSM speed
around 800--1000 km s$^{-1}$; Chugai et al.\ 2004).  One can therefore
also understand the long duration of SN~2006tf in the context of its
CSM speed: the terminal speed of 2000 km s$^{-1}$ reached by the
post-shock shell of SN~2006tf is still much greater than the 190 km
s$^{-1}$ speed of its CSM.  In other SNe~IIn with faster CSM, one
might expect the interaction luminosity to drop as the shock
decelerates, because the smaller difference in speed yields a weaker
shock.

The mass-loss rates of 1--2 M$_{\odot}$ yr$^{-1}$ for about 10 yr
before core collapse, and 0.1--0.2 M$_{\odot}$ yr$^{-1}$ for several
decades before that, are truly astounding.  The only stars known to be
capable of producing it are very massive LBVs during their so-called
``giant eruptions'' (Smith \& Owocki 2006).  Normal stellar winds,
even those of very massive and luminous stars, do not even come close.
Thus, SN~2006tf adds a well established and rather extreme case to the
growing body of evidence requiring massive eruptions in the decades
preceding the core collapse of an initially very massive star.  For
SN~2006tf, the kinetic energy of that ejection was about 10$^{48.8}$
ergs (assuming 18~M$_{\odot}$ moving at 190 km s$^{-1}$).  In both SN
progenitors and LBVs, these eruptions have no accepted explanation.
For the SN progenitors, one potential origin comes from the
pulsational pair instability discussed by Woosley et al.\ (2007) for
SN~2006gy.  That hypothesis, though, requires that the initial masses
of these stars be 100--150 M$_{\odot}$, so any scenario would require
that both SNe~2006gy and 2006tf must have marked the deaths of very
massive stars.

The type of diffusion model we suggest provides an explanation for one
extremely important aspect of SN~2006tf where a traditional
CSM-interaction model fails: the blast-wave speed of SN~2006tf remains
steady at a surprisingly slow value of 2000 km s$^{-1}$, even from
very early times only 32~d after discovery, despite the fact that it
is radiating almost 10$^{51}$ ergs in visual light.  Traditional CSM
interaction predicts an observable deceleration with time (e.g.,
Chugai \& Danziger 1994, 2003), while SNe~1988Z and 1998S provide
observed examples where deceleration is clearly documented (Chugai \&
Danziger 1994; Leonard et al.\ 2000; Pozzo et al.\ 2004).  In the
opaque shell-shocked model that we suggest for SN~2006tf, however, the
bulk of the deceleration occurs very early as the blast wave runs
through the opaque shell.  Delayed photon diffusion allows the energy
to be radiated away later.  The massive circumstellar envelope that
has been accelerated then coasts at a constant speed of 2000 km
s$^{-1}$ (so $R/R_0$ is not large and adiabatic losses are minimized)
while its thermal energy is radiated away as the shell expands, thins,
and cools.

\acknowledgments 
\footnotesize

We thank an anonymous referee for helpful comments.  Some of the data
presented herein were obtained at the W. M. Keck Observatory, which is
operated as a scientific partnership among the California Institute of
Technology, the University of California, and the National Aeronautics
and Space Administration (NASA). The Observatory was made possible by
the generous financial support of the W.M.\ Keck Foundation.  We wish
to extend special thanks to those of Hawaiian ancestry on whose sacred
mountain we are privileged to be guests.  KAIT was constructed and
supported by donations from Sun Microsystems, Inc., the
Hewlett-Packard Company, AutoScope Corporation, Lick Observatory, the
US National Science Foundation (NSF), the University of California,
the Sylvia \& Jim Katzman Foundation, and the TABASGO
Foundation. A.V.F.'s supernova group at U.C. Berkeley is supported by
NSF grant AST--0607485 and by the TABASGO Foundation.  Research by
A.J.B.\ is supported by NSF grant AST--0548198.  We thank Misty Bentz
for assistance with the Keck/ESI observing run.


\end{document}